\documentclass[showpacs,showkeys,twoside,eqsecnum,prd,twocolumn]{revtex4}
\usepackage{amsfonts,amsmath,amssymb,bm,hyperref,graphicx}

\begin{document}

\preprint{E-print archive: 0809.0963 [hep-ph]}

\title{Oscillations of Dirac and Majorana neutrinos in matter and a magnetic field}

\author{Maxim Dvornikov$^{a,b,c}$}
\email{maxim.dvornikov@usm.cl}
\author{Jukka Maalampi$^{a,d}$}
\email{maalampi@cc.jyu.fi} \affiliation{$^a$Department of Physics,
P.O.~Box~35, FIN-40014 University of Jyv\"askyl\"a, Finland;\\
$^b$Departamento de F\'{i}sica y Centro de Estudios
Subat\'{o}micos, Universidad T\'{e}cnica Federico Santa Mar\'{i}a,
Casilla 110-V, Valpara\'{i}so, Chile;\\
$^c$IZMIRAN, 142190, Troitsk, Moscow Region, Russia;\\
$^d$Helsinki Institute of Physics, P.O.~Box~64, FIN-00014\textit{}
University of Helsinki, Finland}

\date{\today}

\begin{abstract}
We study the evolution of massive mixed Dirac and Majorana
neutrinos in matter under the influence of a transversal magnetic
field. The analysis is based on relativistic quantum mechanics. We
solve exactly the evolution equation for relativistic neutrinos,
find the neutrino wave functions, and calculate the transition
probability for spin-flavor oscillations. We analyze the
dependence of the transition probability on the external fields
and compare the cases of Dirac and Majorana neutrinos. The
evolution of Majorana particles in vacuum is also studied and
correction terms to the standard oscillation formula are derived
and discussed. As a possible application of our results we discuss
the spin-flavor transitions in supernovae.
\end{abstract}



\maketitle

\section{Introduction}

If massive neutrinos possess nonzero transition magnetic moments,
transitions $\nu_\beta^\mathrm{L} \leftrightarrow
\nu_\alpha^\mathrm{R}$ that change both spin and flavor of
neutrinos can happen in electromagnetic fields. Such transitions
may be realized, e.g., in astrophysical environments where strong
magnetic fields are present. Oscillations of Dirac neutrinos in an
external magnetic field were studied in Ref.~\cite{DiracBosc}. In
this scenario one has a transition to the sterile neutrino state
$\nu_\mathrm{R}$. While considering oscillations of Majorana
neutrinos with transition magnetic moments in an external magnetic
field (see, e.g., Refs.~\cite{SchVal81,LimMar88,Akh88}), one has
transitions between active neutrino states since
$\nu_\alpha^\mathrm{R}=(\nu_\alpha^\mathrm{L})^c$ for a Majorana
particle.

Neutrino spin-flavor oscillations in solar magnetic fields have
been earlier studied in connection to the solar neutrino problem
(see, e.g., Ref.~\cite{OscSolarNu}). It was thought that this
neutrino oscillations channel could explain at least partly the
deficit of the electron neutrinos in the measured solar neutrino
flux. It is now clear that such spin-flavor oscillations cannot
play any significant role  and the deficit can be satisfactorily
explained in terms of active-active conversions and the
Mikheyev-Smirnov-Wolfenstein matter effect~\cite{smallcontr}.

The influence of strong magnetic fields of neutron stars on
neutrino oscillations was studied in
Refs.~\cite{OscSNNu,AndSatPRD03,AndSatJCAP03}. It was shown that
spin-flavor transition may  have important effects in the neutron
star environment. In these investigations neutrino spin-flavor
oscillations were described for  realistic profiles of matter
densities and magnetic fields and the  appearance of resonances in
neutrino oscillations was examined.  For further details on the
neutrino oscillations in electromagnetic fields as well as
neutrino electromagnetic properties the reader is referred to the
recent review~\cite{GiuStu08}.

In this paper we shall return to the question of the spin-flavor
oscillations using an approach that differs from the one usually
followed. It is usual to describe neutrino oscillations, including
the spin-flavor transitions, on the basis of the quantum
mechanical evolution equation. Instead of this Schr\"odinger
picture, we will apply relativistic quantum mechanical picture
based on the Dirac theory. In this approach, we will  study
spin-flavor oscillations of Dirac and Majorana neutrinos in matter
and in an external magnetic field, extending our earlier use of
the method~\cite{FOvac,Dvo06EPJC,DvoMaa07,Dvo08} to this new
problem. We should also mention that the majority of the previous
studies of neutrino spin-flavor oscillations are restricted to the
case of Majorana neutrinos. We will investigate both the Dirac and
Majorana cases as the nature of neutrinos is still an open
question~\cite{EllEng04}.


The plan of this paper is as follows. We will start by writing
down the relativistic wave equations that take into account the
background matter and external magnetic field. We formulate the
initial condition problem for these systems (see also
Refs.~\cite{FOvac,Dvo06EPJC,DvoMaa07,Dvo08}). We then derive for
both Dirac and Majorana neutrinos a Hamiltonian analogous to that
of the standard quantum mechanical approach and solve exactly  the
resulting evolution equation for relativistic neutrinos. We then
analyze the behavior of the transition probability for Dirac and
Majorana  neutrinos at various magnetic field strengths and matter
densities. Our results will be summarized and their applications
to some astrophysical situations  discussed  in Sec.~\ref{CONCL}.


\clearpage

\section{Evolution of Dirac neutrinos in matter and transversal magnetic
field}\label{DIRNUMATTB}

Let us study the evolution of two neutrino flavor states
$\nu_{\lambda}$, $\lambda=\alpha,\beta$, in a nonmoving and
unpolarized matter under the influence of an external magnetic
field. We assume that the mass eigenstates $\psi_a$, $a=1,2$, are
related to the flavor eigenstate neutrinos through the
transformation
\begin{equation}\label{matrtransDir}
  \nu_\lambda=\sum_{a=1,2}U_{\lambda a}\psi_a,
  \quad
  ({U}_{\lambda a})=
  \begin{pmatrix}
    \cos \theta & -\sin \theta \\
    \sin \theta & \cos \theta \\
  \end{pmatrix},
\end{equation}
where $\theta$ is a mixing angle.  In this section we shall assume
that the mass eigenstates are Dirac particles. We set the
following initial conditions (see also
Refs.~\cite{FOvac,Dvo06EPJC,DvoMaa07,Dvo08}):
\begin{equation}\label{IniCondDir}
  \nu_\alpha(\mathbf{r},0)=0,
  \quad
  \nu_\beta(\mathbf{r},0)=\nu_\beta^{(0)}e^{\mathrm{i}\mathbf{k}\mathbf{r}},
\end{equation}
where $\mathbf{k}=(k,0,0)$ is the initial momentum and
$\nu_\beta^{(0)\mathrm{T}} = (1/2)(1,-1,-1,1)$ (see below). This
corresponds to a relativistic neutrino of the flavor $\beta$, with
its spin directed oppositely to the particle momentum, i.e. a
left-handed neutrino $\nu_\beta^\mathrm{L}$. The system is taken
not to contain the other neutrino flavor $\nu_{\alpha}$ initially.

Note that from the physical point of view it would be more
realistic to choose a localized in space initial condition rather
than that in Eq.~\eqref{IniCondDir}. It is well known, however,
that from the point of view of oscillations the plane wave and
wave packet approaches in practice lead to equivalent
results~\cite{Giunti:2007ry}. Moreover the initial condition
problem~\eqref{IniCondDir} for the cases of flavor and spin-flavor
oscillations in various external fields was solved in our previous
publications~\cite{FOvac,Dvo06EPJC,DvoMaa07,Dvo08}.

The Dirac equation for the mass eigenstates wave functions is of
the form (see Refs.~\cite{DvoMaa07,Dvo08}),
\begin{equation}\label{WaveEqDirNu}
  \mathrm{i}\dot{\psi}_a = \mathcal{H}_a \psi_a +(V_m + V_B) \psi_b,
  \quad
  a \neq b,
\end{equation}
where $\mathcal{H}_a = (\bm{\alpha}\mathbf{p})+\beta m_a + g_a
(1-\gamma_5)/2$ is the mass-diagonal part of the Hamiltonian,
$m_a$ are the masses associated with the states $\psi_a$, and
$\bm{\alpha}=\gamma^0\bm{\gamma}$, $\beta=\gamma^0$, and
$\bm{\Sigma}=\gamma^5\gamma^0\bm{\gamma}$ are Dirac matrices. The
term $V_m = g (1-\gamma_5)/2$ describes the interaction of
neutrinos with particles in the background matter. The matrix
$(g_{ab})$, given below,  is not diagonal in the mass eigenstate
basis. The term $V_B = -\mu B \beta \Sigma_3$ is the energy
operator associated with the interactions of neutrinos with the
magnetic field, and the magnetic moment matrix $(\mu_{ab})$ is,
like $(g_{ab})$, in general nondiagonal. The magnetic field is
taken to be transversal with respect to the initial neutrino
momentum, i.e., $\mathbf{B} = (0,0,B)$. The $V_m$ and $V_B$ terms
are responsible for the possible  mixing between different
neutrino mass eigenstates.

The matrix $(g_{ab})$ that describes the interactions of neutrinos
with matter is in the mass eigenstates basis of the form
\begin{align}\label{gexpl}
  g_{ab} = &
  \sum_{\lambda=\alpha\beta}
  U_{a\lambda}^\dag f_\lambda U_{\lambda b}
  \\
  \notag
  = &
  \begin{pmatrix}
    f_\alpha\cos^2 \theta + f_\beta\sin^2 \theta & -\sin\theta\cos\theta[f_\alpha-f_\beta] \\
    -\sin\theta\cos\theta[f_\alpha-f_\beta] & f_\alpha \sin^2\theta + f_\beta \cos^2\theta \\
  \end{pmatrix}.
\end{align}
We will denote $g_{aa}=g_a$ and $g_{12}=g_{21}=g$. If we identify
the flavor $\alpha$ as $\equiv \nu_\mu$ or $\nu_\tau$ and the
flavor $\beta$ as $\equiv \nu_e$, the effective potentials
$f_\lambda$, $\lambda=\alpha,\beta$, are given by (see
Ref.~\cite{matteffpot}):
\begin{align}\label{flambda}
  f_\lambda = & \sqrt{2}G_\mathrm{F}
  \sum_{f=e,p,n} n_f q_f^{(\lambda)},
  \notag
  \\
  q_f^{(\alpha)} = & (I_{3\mathrm{L}}^{(f)}-2Q^{(f)}\sin^2\theta_W),
  \notag
  \\
  q_f^{(\beta)} = & (I_{3\mathrm{L}}^{(f)}-2Q^{(f)}\sin^2\theta_W+\delta_{ef}),
\end{align}
where $n_f$ is the number density, $I_{3\mathrm{L}}^{(f)}$  the
third isospin component and $Q^{(f)}$ the electric charge of the
background particle of the type $f$, $\theta_{W}$ is the weak
mixing angle, and $G_\mathrm{F}$ is the Fermi constant. It is
assumed in  Eq.~\eqref{flambda}  that matter consists of
electrons, protons, and neutrons and that it is unpolarized and at
rest.

As to the interactions of neutrinos with the magnetic field, we
have assumed when writing Eq.~\eqref{WaveEqDirNu} that the
magnetic moment matrix in the mass eigenstates basis is
antidiagonal, i.e., $\mu_{aa}=0$ and $\mu_{12}=\mu_{21}=\mu \neq
0$. Magnetic moment matrices of this type were studied in our
previous works~\cite{DvoMaa07,Dvo08}. The situation with the
values of Dirac neutrino magnetic moments is
disputable~\cite{Bel05}; however, the existence of large
off-diagonal magnetic moments of Dirac neutrinos is not excluded.
Let us note that the solution to the Dirac-Pauli equation for a
neutrino with $\mu_{aa} \neq 0$ propagating in an arbitrary moving
and polarized medium was recently obtained in
Ref.~\cite{ArbLobMur07}. However in that work the case of only one
neutrino flavor was studied.

The general solution of Eq.~\eqref{WaveEqDirNu} has the
form~\cite{Dvo06EPJC,DvoMaa07,Dvo08},
\begin{align}\label{GenSolDirNu}
  \psi_{a}(\mathbf{r},t)= &
  \exp{(-\mathrm{i} g_a t/2)}
  \int \frac{\mathrm{d}^3 \mathbf{p}}{(2\pi)^{3/2}}
  e^{\mathrm{i} \mathbf{p} \mathbf{r}}
  \notag
  \\
  & \times
  \sum_{\zeta=\pm 1}
  \Big[
    a_a^{(\zeta)}(t)u_a^{(\zeta)}\exp{(-\mathrm{i}E_a^{(\zeta)} t)}
    \notag
    \\
    & +
    b_a^{(\zeta)}(t)v_a^{(\zeta)}\exp{(+\mathrm{i}E_a^{(\zeta)} t)}
  \Big],
\end{align}
where the energies are given by (see, e.g., Ref.~\cite{matterQFT})
\begin{equation}\label{EnergyLevQFT}
  E_a^{(\zeta)}=\sqrt{m_a^2 + (|\mathbf{p}| - \zeta g_a/2)^2}.
\end{equation}
In the relativistic limit one has
\begin{equation}\label{relEnergyLevQFT}
  E_a^{(\zeta)} \approx
  |\mathbf{p}| + \frac{m_a^2}{2|\mathbf{p}|} -\zeta \frac{g_a}{2}.
\end{equation}
To obtain Eq.~\eqref{relEnergyLevQFT} we neglect the term $\sim
g_a^2$ compared to the neutrino mass squared $m_a^2$. For the
situation of neutrino propagation in the expanding envelope
formed after the supernova explosion, which is studied below, the
effective potentials are $g_a \sim 10^{-12} -
10^{-11}\thinspace\text{eV}$. Accounting for the neutrino mass in
the $\text{eV}$ range, the approximation made is always valid.

The expressions for the basis spinors $u_a^{(\zeta)}$ and
$v_a^{(\zeta)}$ can be found in Ref.~\cite{matterQFT}. When one
studies the situation where the initial momentum is very large, $k
\gg m_a$, one can neglect the neutrino mass dependence of the
spinors and present them in the form
\begin{align}\label{BspinorsDir}
  u^{+{}} = &
  \frac{1}{2}
  \begin{pmatrix}
    1 \\
    1 \\
    1 \\
    1
  \end{pmatrix},
  \quad
  u^{-{}} =
  \frac{1}{2}
  \begin{pmatrix}
    1 \\
    -1 \\
    -1 \\
    1
  \end{pmatrix},
  \notag
  \\
  v^{+{}} = &
  \frac{1}{2}
  \begin{pmatrix}
    1 \\
    1 \\
    -1 \\
    -1
  \end{pmatrix},
  \quad
  v^{-{}} =
  \frac{1}{2}
  \begin{pmatrix}
    1 \\
    -1 \\
    1 \\
    -1
  \end{pmatrix}.
\end{align}
Note that the coefficients $a_a^{(\zeta)}(t)$ and
$b_a^{(\zeta)}(t)$ in Eq.~\eqref{GenSolDirNu} are functions of
time due to the presence of the nondiagonal interactions $V_m$ and
$V_B$. Our main goal is to determine these coefficients so that
both Eq.~\eqref{WaveEqDirNu} and the initial
condition~\eqref{IniCondDir} are satisfied.

Using Eqs.~\eqref{WaveEqDirNu} and \eqref{GenSolDirNu} and the
orthonormality of the basis spinors~\eqref{BspinorsDir} we arrive
to the following ordinary differential equations for the
coefficients $a_a^{(\zeta)}$ and
$b_a^{(\zeta)}$~\cite{DvoMaa07,Dvo08}:
\begin{align}\label{ODEDir}
  \mathrm{i}\dot{a}_a^{(\zeta)}= &
  \exp{[\mathrm{i}(g_a-g_b)t/2]}
  \exp{(+\mathrm{i}E_a^{(\zeta)}t)}u^{(\zeta)\dag}
  (V_m + V_B)
  \notag
  \\
  & \times
  \sum_{\zeta'=\pm 1}
  \Big[
    a_b^{(\zeta')}u^{(\zeta')}\exp{(-\mathrm{i}E_b^{(\zeta')} t)}
    \notag
    \\
    & +
    b_b^{(\zeta')}v^{(\zeta')}\exp{(+\mathrm{i}E_b^{(\zeta')} t)}
  \Big],
  \notag
  \\
  \mathrm{i}\dot{b}_a^{(\zeta)}= &
  \exp{[\mathrm{i}(g_a-g_b)t/2]}
  \exp{(-\mathrm{i}E_a^{(\zeta)} t)}v^{(\zeta)\dag}
  (V_m + V_B)
  \notag
  \\
  & \times
  \sum_{\zeta'=\pm 1}
  \Big[
    a_b^{(\zeta')}u^{(\zeta')}\exp{(-\mathrm{i}E_b^{(\zeta')} t)}
    \notag
    \\
    & +
    b_b^{(\zeta')}v^{(\zeta')}\exp{(+\mathrm{i}E_b^{(\zeta')} t)}
  \Big].
\end{align}
One easily sees that $\langle u^{-{}} | V_m | u^{-{}} \rangle =
g$, $\langle v^{+{}} | V_m | v^{+{}} \rangle = g$, $\langle
u^{\pm{}} | V_B | u^{\mp{}} \rangle = -\mu B$ and $\langle
v^{\pm{}} | V_B | u^{\mp{}} \rangle = -\mu B$. All the other
scalar products in Eq.~\eqref{ODEDir} vanish.

Let us introduce a four-component wave function $\Psi'$ defined as
$\Psi^{'\mathrm{T}}=(a_1^{-{}},a_2^{-{}},a_1^{+{}},a_2^{+{}})$.
Then we can rewrite Eq.~\eqref{ODEDir} in the form of a
Schr\"odinger equation:
\begin{widetext}
\begin{equation}\label{QMH1Dir}
  \mathrm{i}\frac{\mathrm{d}\Psi'}{\mathrm{d}t} = H' \Psi',
  \quad
  H' =
  \begin{pmatrix}
    0 & g e^{\mathrm{i} \omega' t} & 0 & -\mu B e^{\mathrm{i} \omega_1 t} \\
    g e^{-\mathrm{i} \omega' t} & 0 & -\mu B e^{-\mathrm{i} \omega_2 t} & 0 \\
    0 & -\mu B e^{\mathrm{i} \omega_2 t} & 0 & 0 \\
    -\mu B e^{-\mathrm{i} \omega_1 t} & 0 & 0 & 0 \
  \end{pmatrix},
\end{equation}
%
where [see also Eq.~\eqref{relEnergyLevQFT}]
\begin{gather}
  \omega_1  = E_1^{-{}}-E_2^{+{}} + \frac{g_1-g_2}{2} \approx
  2 \Phi + g_1,
  \quad
  \omega_2 = E_1^{+{}}-E_2^{-{}} + \frac{g_1-g_2}{2} \approx
  2 \Phi - g_2,
  \notag
  \\
  \label{omegaDefDir}
  \omega' = E_1^{-{}}-E_2^{-{}} + \frac{g_1-g_2}{2} \approx
  2 \Phi + 2(g_1 - g_2),
\end{gather}
and $\Phi = \delta m^2/(4k)=(m_1^2-m_2^2)/(4k)$ is the vacuum
oscillation phase.

Let us next do the following transformation to the wave function:
%
\begin{align*}
  \Psi' = & \mathcal{U}\Psi,
  \\
  \mathcal{U} = & \mathrm{diag}
  \left\{
    e^{\mathrm{i}(\Phi+3g_1/4-g_2/4)t},
    e^{-\mathrm{i}(\Phi-3g_2/4+g_1/4)t},
    e^{\mathrm{i}(\Phi-g_1/4-g_2/4)t},
    e^{-\mathrm{i}(\Phi+g_1/4+g_2/4)t}
  \right\}.
\end{align*}
%
This takes  the Schr\"odinger Eq.~\eqref{QMH1Dir} into the form
%
\begin{align}\label{QMH2Dir}
  \mathrm{i}\frac{\mathrm{d}\Psi}{\mathrm{d}t} = & H \Psi,
  \quad
  H =
  \mathcal{U}^\dag H' \mathcal{U} -
  \mathrm{i} \mathcal{U}^\dag \dot{\mathcal{U}} =
  \\
  \notag
  = &
  \begin{pmatrix}
    \Phi + 3g_1/4 - g_2/4 & g & 0 & -\mu B \\
    g & -\Phi + 3g_2/4 - g_1/4 & -\mu B & 0 \\
    0 & -\mu B & \Phi - (g_1+g_2)/4 & 0 \\
    -\mu B & 0 & 0 & -\Phi - (g_1+g_2)/4 \
  \end{pmatrix}.
\end{align}
\end{widetext}

It is in order at this stage to compare the time evolution
Eq.~\eqref{QMH2Dir}, which we obtained to in our approach based on
Dirac equation, with the one obtained in the conventional quantum
mechanical formalism. To this end let us introduce a quantum
mechanical wave function $\Psi_{QM}^\mathrm{T} =
(\psi_1^\mathrm{L}, \psi_2^\mathrm{L}, \psi_1^\mathrm{R},
\psi_2^\mathrm{R})$ for the neutrino mass eigenstates. Here
$\psi_a^\mathrm{L,R}$ are one-component objects. This wave
function obeys the Schr\"odinger equation with the following
effective Hamiltonian (see, e.g., Ref.~\cite{LimMar88}):
\begin{equation}\label{QMH3Dir}
  H_{QM} =
  \begin{pmatrix}
    \mathcal{K}_1 + g_1 & g & 0 & -\mu B \\
    g & \mathcal{K}_2 + g_2 & -\mu B & 0 \\
    0 & -\mu B & \mathcal{K}_1 & 0 \\
    -\mu B & 0 & 0 & \mathcal{K}_2 \
  \end{pmatrix},
\end{equation}
where $\mathcal{K}_a = \sqrt{k^2+m_a^2} \approx k + m_a^2/(2k)$ is
the kinetic energy of a massive neutrino. It is easy to see that
this matrix leads  to the same dynamics as the Hamiltonian in
Eq.~\eqref{QMH2Dir}. Indeed, without changing the dynamics one can
make the following replacement $H_{QM} \to H_{QM} -
\mathrm{tr}(H_{QM})/4 \cdot I$, where $I$ is the $4 \times 4$ unit
matrix. The resulting Hamiltonian is exactly the same Hamiltonian
we have derived in our approach.

To describe the evolution of the system~\eqref{QMH2Dir} in a
general case, one has to solve a secular equation which is the
fourth-order algebraic equation (see also below). Although one can
express the solution to such an equation in radicals, its actual
form appears to be rather cumbersome for arbitrary parameters. If
we, however, consider the case of a neutrino propagating in the
electrically neutral isoscalar matter, i.e. $n_e = n_p$ and $n_p =
n_n$, a reasonable solution is possible to find. We will
demonstrate later that it corresponds to a realistic physical
situation. As one can infer from Eq.~\eqref{flambda} for the case
of the $\nu_e^\mathrm{L}\to\nu_\mu^\mathrm{R}$ oscillations
channel, in a medium with this property one has the effective
potentials $f_\alpha \equiv f_\mu = V_\mu = -G_\mathrm{F}
n/\sqrt{2}$ and $f_\beta \equiv f_e = V_e = G_\mathrm{F}
n/\sqrt{2}$, where $n\equiv n_e = n_p = n_n$. Using
Eq.~\eqref{gexpl} we obtain that $g_1 = -g_2 = g_0$, where $g_0 =
-V\cos 2\theta$, $g=V\sin 2\theta$, and $V  =
G_\mathrm{F}n/\sqrt{2}$.

Let us point out that background matter with these properties may
well exist in some astrophysical environments. The matter profile
of presupernovae is poorly known, and a variety of presupernova
models with different profiles exist in the literature (see, e.g.
Ref.~\cite{Woosley:2002zz}). Nevertheless, electrically neutral
isoscalar matter may well exist in the inner parts of
presupernovae consisting of elements heavier than hydrogen.
Indeed, for example, the model W02Z in Ref.~\cite{Woosley:2002zz}
predicts that in a 15$M_{\odot}$ presupernova one has
$Y_e=n_e/(n_p+n_n)=0.5$ in the O+Ne+Mg layer, between the Si+O and
He layers, in the radius range (0.007--0.2)$R_{\odot}$.

In this kind of background matter the effective Hamiltonian in
Eq.~\eqref{QMH2Dir} is replaced by
\begin{equation}\label{QMH4Dir}
  H =
  \begin{pmatrix}
    \Phi+g_0 & g & 0 & -\mu B \\
    g & -(\Phi+g_0) & -\mu B & 0 \\
    0 & -\mu B & \Phi & 0 \\
    -\mu B & 0 & 0 & -\Phi \
  \end{pmatrix}.
\end{equation}
We now look for the stationary solutions of the Schr\"odinger
equation with this Hamiltonian. After a straightforward
calculation one finds
\begin{align}\label{WFDir}
  \Psi(t)= & \sum_{\zeta = \pm 1}
  \Big[
    \left(
      U_\zeta \otimes U_\zeta^\dag
    \right)\exp{(-\mathrm{i}\mathcal{E}_\zeta t)}
    \notag
    \\
    & +
    \left(
      V_\zeta \otimes V_\zeta^\dag
    \right)\exp{(\mathrm{i}\mathcal{E}_\zeta t)}
  \Big]\Psi_0,
\end{align}
where we have denoted
\begin{align}\label{EnergyQMDir}
  \mathcal{E}_{\pm{}}= & \frac{1}{2}
  \sqrt{2V^2+4(\mu B)^2+4\Phi^2-4\Phi V\cos 2\theta \pm 2VR},
  \notag
  \\
  R = & \sqrt{(V-2\Phi\cos 2\theta)^2+4(\mu B)^2}.
\end{align}
The vectors $U_{\pm{}}$ and $V_{\pm{}}$ are the eigenvectors
corresponding to the energy eigenvalues $\mathcal{E}_{\pm{}}$ and
$-\mathcal{E}_{\pm{}}$, respectively. They are given by ($\zeta
=\pm{}$)
\begin{align}\label{UVQMspinorsDir}
  U_\zeta = &
  \frac{1}{N_\zeta}
  \begin{pmatrix}
    Z_\zeta \\
    \sin 2\theta (\mathcal{E}_\zeta-\Phi) \\
    -\mu B \sin 2\theta \\
    -\mu B Z_\zeta/(\mathcal{E}_\zeta + \Phi) \
  \end{pmatrix},
  \notag
  \\
  V_\zeta = &
  \frac{1}{N_\zeta}
  \begin{pmatrix}
    -\sin 2\theta (\mathcal{E}_\zeta-\Phi) \\
    Z_\zeta \\
    \mu B Z_\zeta/(\mathcal{E}_\zeta + \Phi) \\
    -\mu B \sin 2\theta \
  \end{pmatrix},
\end{align}
where
\begin{align}\label{ZNDir}
  Z_\zeta = & \frac{V + \zeta R}{2}-\mathcal{E}_\zeta \cos 2\theta,
  \notag
  \\
  N_\zeta^2 = & Z_\zeta^2
  \left[
    1+\frac{(\mu B)^2}{(\mathcal{E}_\zeta+\Phi)^2}
  \right]
  \notag
  \\
  & +
  \sin^2(2\theta)
  \left[
    (\mu B)^2+(\mathcal{E}_\zeta-\Phi)^2
  \right].
\end{align}
It should be noted that Eq.~\eqref{WFDir} is the general solution
of Eq.~\eqref{QMH2Dir} satisfying the initial condition
$\Psi(0)=\Psi_0$.

Note that we received the solution~\eqref{WFDir}-\eqref{ZNDir} of
the evolution Eq.~\eqref{QMH2Dir} under some assumptions on the
external fields such as isoscalar matter with constant density and
constant magnetic field. We mentioned above that our method is
equivalent to the quantum mechanical description of neutrino
oscillations~\cite{LimMar88} which is valid for a more general
case of coordinate dependent external fields. The advantage of our
approach consists in the fact that one can derive neutrinos' wave
functions for arbitrary initial momenta, as it was made in
Refs.~\cite{FOvac,Dvo06EPJC}, and study the propagation of
low-energy neutrinos. The assumption of constant matter density
and magnetic field is quite realistic for certain astrophysical
environments like a shock wave propagating inside an expanding
envelope after a supernova explosion (see also Sec.~\ref{CONCL}).

Consistently with Eqs.~\eqref{matrtransDir}
and~\eqref{IniCondDir}, we take the initial wave function
$\Psi(0)\equiv\Psi_0$ in Eq.~\eqref{WFDir} as
$\Psi_0^\mathrm{T}=(\psi_1^\mathrm{L}, \psi_2^\mathrm{L},
\psi_1^\mathrm{R},
\psi_2^\mathrm{R})=(\sin\theta,\cos\theta,0,0)$. Using
Eqs.~\eqref{WFDir}-\eqref{ZNDir} one finds the components of the
quantum mechanical wave function corresponding to the right-handed
neutrinos to be of the form
\begin{align}\label{psi1Rpsi2RDir}
  \psi_1^\mathrm{R}(t) = &
  \frac{\mu B}{N_{+{}}^2}
  \bigg\{
    \cos\theta
    \bigg[
      e^{\mathrm{i}\mathcal{E}_{+{}}t}
      \frac{Z_{+{}}^2}{\mathcal{E}_{+{}}+\Phi}
      \notag
      \\
      & -
      \sin^2(2\theta)(\mathcal{E}_{+{}}-\Phi)e^{-\mathrm{i}\mathcal{E}_{+{}}t}
    \bigg]
    \notag
    \\
    & -
    \sin \theta \sin 2\theta Z_{+{}}
    \left[
      e^{-\mathrm{i}\mathcal{E}_{+{}}t}+
      e^{\mathrm{i}\mathcal{E}_{+{}}t}
      \frac{\mathcal{E}_{+{}}-\Phi}{\mathcal{E}_{+{}}+\Phi}
    \right]
  \bigg\}
  \notag
  \\
  & +
  \{+{}\to{}-{}\},
  \notag
  \\
  \psi_2^\mathrm{R}(t) = &
  \frac{\mu B}{N_{+{}}^2}
  \bigg\{
    \sin\theta
    \bigg[
      \sin^2(2\theta)(\mathcal{E}_{+{}}-\Phi)e^{\mathrm{i}\mathcal{E}_{+{}}t}
      \notag
      \\
      & -
      e^{-\mathrm{i}\mathcal{E}_{+{}}t}
      \frac{Z_{+{}}^2}{\mathcal{E}_{+{}}+\Phi}
    \bigg]
    \notag
    \\
    & -
    \cos \theta \sin 2\theta Z_{+{}}
    \left[
      e^{\mathrm{i}\mathcal{E}_{+{}}t}+
      e^{-\mathrm{i}\mathcal{E}_{+{}}t}
      \frac{\mathcal{E}_{+{}}-\Phi}{\mathcal{E}_{+{}}+\Phi}
    \right]
  \bigg\}
  \notag
  \\
  & +
  \{+{}\to{}-{}\},
\end{align}
where the $\{+{}\to{}-{}\}$ stand for the terms similar to the
terms preceding each of them but with all quantities with a
subscript $+{}$ replaced with corresponding quantities with a
subscript $-{}$. The wave function of the right-handed neutrino of
the flavor $\alpha$, $\nu_\alpha^\mathrm{R}$, can be written with
help of Eqs.~\eqref{matrtransDir} and~\eqref{psi1Rpsi2RDir} as
$\nu_\alpha^\mathrm{R}(t) = \cos\theta\psi_1^\mathrm{R}(t) -
\sin\theta\psi_2^\mathrm{R}(t)$.

The probability for the transition $\nu_\beta^\mathrm{L} \to
\nu_\alpha^\mathrm{R}$ is obtained as the square of the quantum
mechanical wave function $\nu_\alpha^\mathrm{R}$. One obtains
\begin{align}\label{PrtDir}
  P_{\nu_\beta^\mathrm{L} \to \nu_\alpha^\mathrm{R}}(t) = &
  \left|
    \nu_\alpha^\mathrm{R}
  \right|^2=
  [C_{+{}}\cos(\mathcal{E}_{+{}}t)+C_{-{}}\cos(\mathcal{E}_{-{}}t)]^2
  \notag
  \\
  & +
  [S_{+{}}\sin(\mathcal{E}_{+{}}t)+S_{-{}}\sin(\mathcal{E}_{-{}}t)]^2,
\end{align}
where ($\zeta =\pm$)
\begin{align}\label{CpmSpmDir}
  C_\zeta = & \frac{\mu B}{N_\zeta^2}
  \left\{
    \frac{Z_\zeta^2}{\mathcal{E}_\zeta+\Phi}-
    \sin^2(2\theta)(\mathcal{E}_\zeta-\Phi)
  \right\},
  \notag
  \\
  S_\zeta = & \frac{\mu B}{N_\zeta^2}
  \bigg\{
    \sin^2(2\theta)\frac{2\Phi Z_\zeta}{\mathcal{E}_\zeta+\Phi}
    \notag
    \\
    & +
    \cos 2\theta
    \left[
      \frac{Z_\zeta^2}{\mathcal{E}_\zeta+\Phi}+
      \sin^2(2\theta)(\mathcal{E}_\zeta-\Phi)
    \right]
  \bigg\}.
\end{align}
As a consistency check, one easily finds from
Eq.~\eqref{CpmSpmDir} that $C_{+{}}+C_{-{}}=0$ as required for
assuring $P(0)=0$.

In the following we will limit our considerations to the case
 $\mathcal{E}_{+{}} \approx \mathcal{E}_{-{}}$, corresponding
to the situations where the  effect of the interactions of
neutrinos with matter ($V$) is small compared with that of the
magnetic interactions ($\mu B$) or the vacuum contribution
($\Phi$) or both [see Eq.~\eqref{EnergyQMDir}]. Note that in this
case one can analyze the exact oscillation
probability~\eqref{PrtDir} analytically, which would be
practically impossible in more general situations.

In the case  $\mathcal{E}_{+{}} \approx \mathcal{E}_{-{}}$, one
can present the transition probability in Eq.~\eqref{PrtDir} in
the following form:
\begin{equation}\label{PrtEnvDir}
  P(t) = P_0(t) + P_c(t) \cos(2 \Omega t) + P_s(t) \sin(2 \Omega t),
\end{equation}
where
\begin{align}\label{P0PcPsDir}
  P_0(t) = & \frac{1}{2}
  \big[
    S_{+{}}^2+S_{-{}}^2+2 S_{+{}} S_{-{}} \cos(2 \delta \Omega t)
    \notag
    \\
    & -
    4 C_{+{}} C_{-{}} \sin^2(\delta \Omega t)
  \big],
  \notag
  \\
  P_c(t) = & -\frac{1}{2}
  \big[
    (S_{+{}}^2+S_{-{}}^2) \cos(2 \delta \Omega t) + 2 S_{+{}} S_{-{}}
    \notag
    \\
    & -
    4 C_{+{}} C_{-{}} \sin^2(\delta \Omega t)
  \big],
  \notag
  \\
  P_s(t) = & \frac{1}{2}
  \left(
    S_{+{}}^2-S_{-{}}^2
  \right)
  \sin(2 \delta \Omega t),
\end{align}
and
\begin{equation}\label{FrDir}
  \Omega = \frac{\mathcal{E}_{+{}}+\mathcal{E}_{-{}}}{2},
  \quad
  \delta \Omega = \frac{\mathcal{E}_{+{}}-\mathcal{E}_{-{}}}{2}.
\end{equation}
As one can infer from these expressions, the transition
probability $P(t)$ is a rapidly oscillating function,  with the
frequency $\Omega$, enveloped from up and down by the slowly
varying functions $P_{u,d} = P_0 \pm \sqrt{P_c^2+P_s^2}$,
respectively.

\begin{figure*}
  \centering
  \includegraphics[scale=.93]{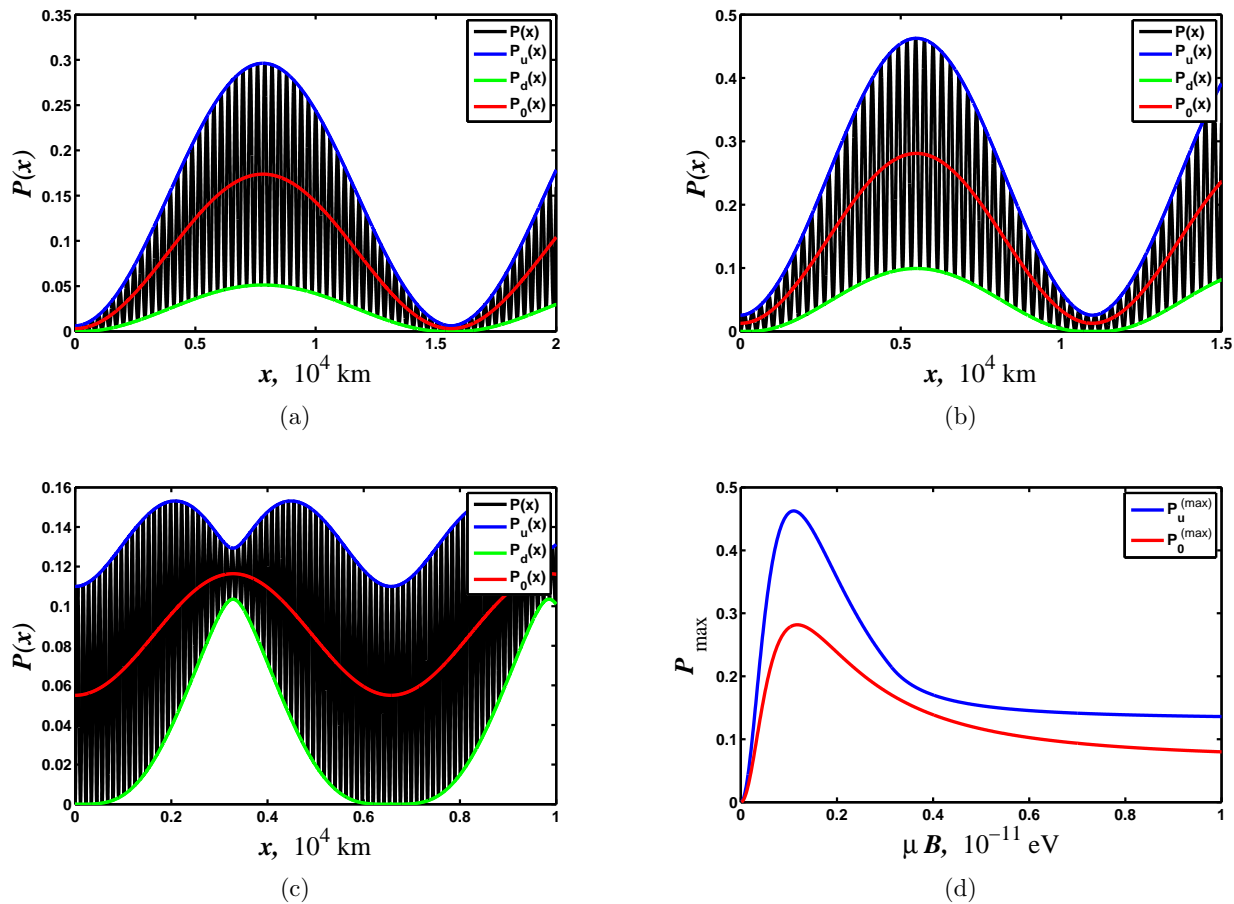}
  \caption{\label{fig1}
  (a)-(c) The transition probability versus the distance passed by a neutrino
  beam in matter with the density $\rho = 10\thinspace\text{g/cc}$;
  (a) $\mu B = 5 \times 10^{-13}\thinspace\text{eV}$,
  (b) $\mu B = 1.1 \times 10^{-12}\thinspace\text{eV}$,
  (c) $\mu B = 5 \times 10^{-12}\thinspace\text{eV}$.
  We take that $E_\nu = 10\thinspace\text{MeV}$,
  $\delta m^2 = 8 \times 10^{-5}\thinspace\text{eV}^2$ and
  $\theta = 0.6$, which is quite close to the solar neutrinos'
  oscillations parameters.
  The black line is the function $P(x)$,
  the blue and green lines are the envelope functions $P_{u,d}(x)$, and
  the red line is the averaged transition probability $P_0(x)$.
  (d) The dependence of the maximal values of the functions $P(x)$ and
  $P_0(x)$, blue and red lines, respectively, on the magnetic energy $\mu B$
  for the given density.}
\end{figure*}

\begin{figure*}
  \centering
  \includegraphics[scale=.93]{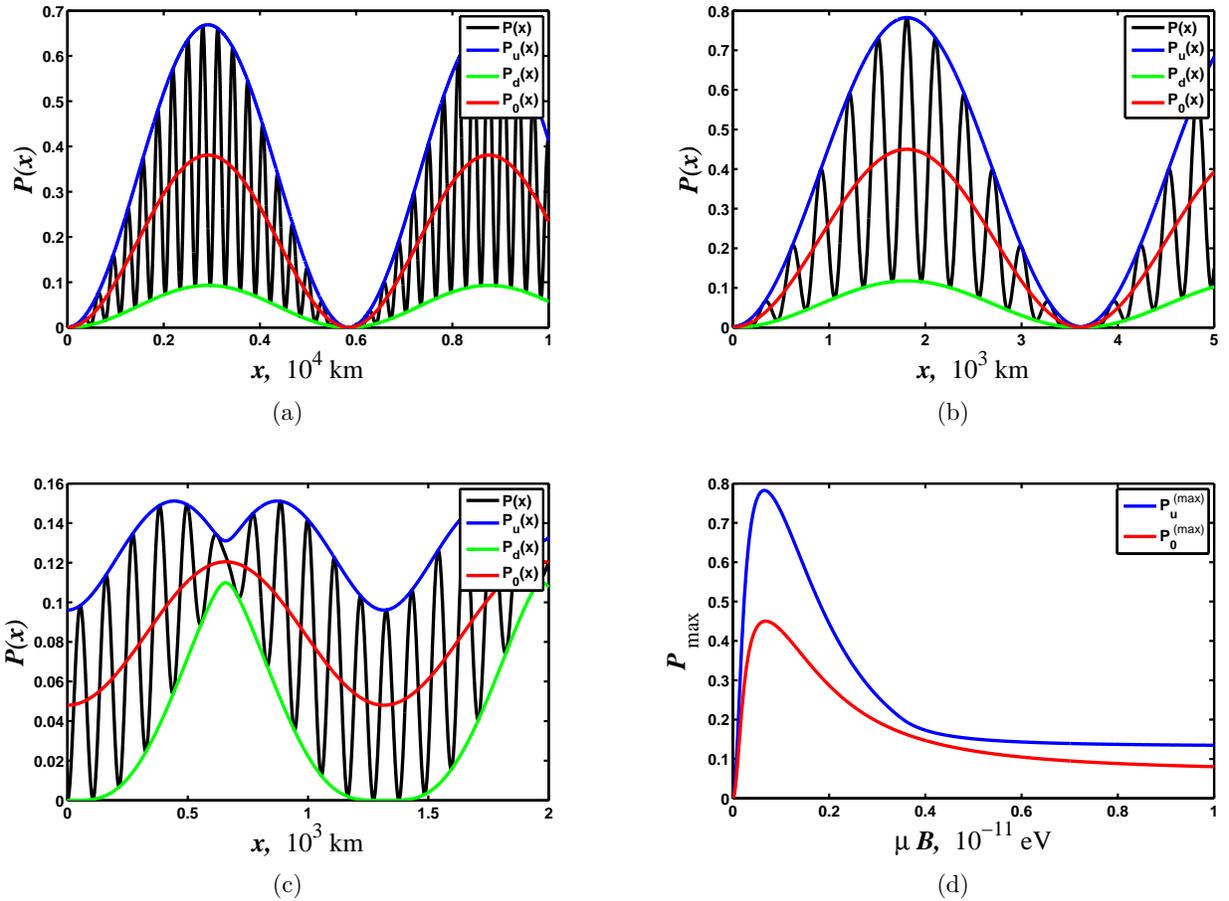}
  \caption{\label{fig2}
  The same as in Fig.~\ref{fig1} for the density
  $\rho = 50\thinspace\text{g/cc}$;
  (a) $\mu B = 3.5 \times 10^{-13}\thinspace\text{eV}$,
  (b) $\mu B = 6.6 \times 10^{-13}\thinspace\text{eV}$,
  (c) $\mu B = 5 \times 10^{-12}\thinspace\text{eV}$.}
\end{figure*}

\begin{figure*}
  \centering
  \includegraphics[scale=.93]{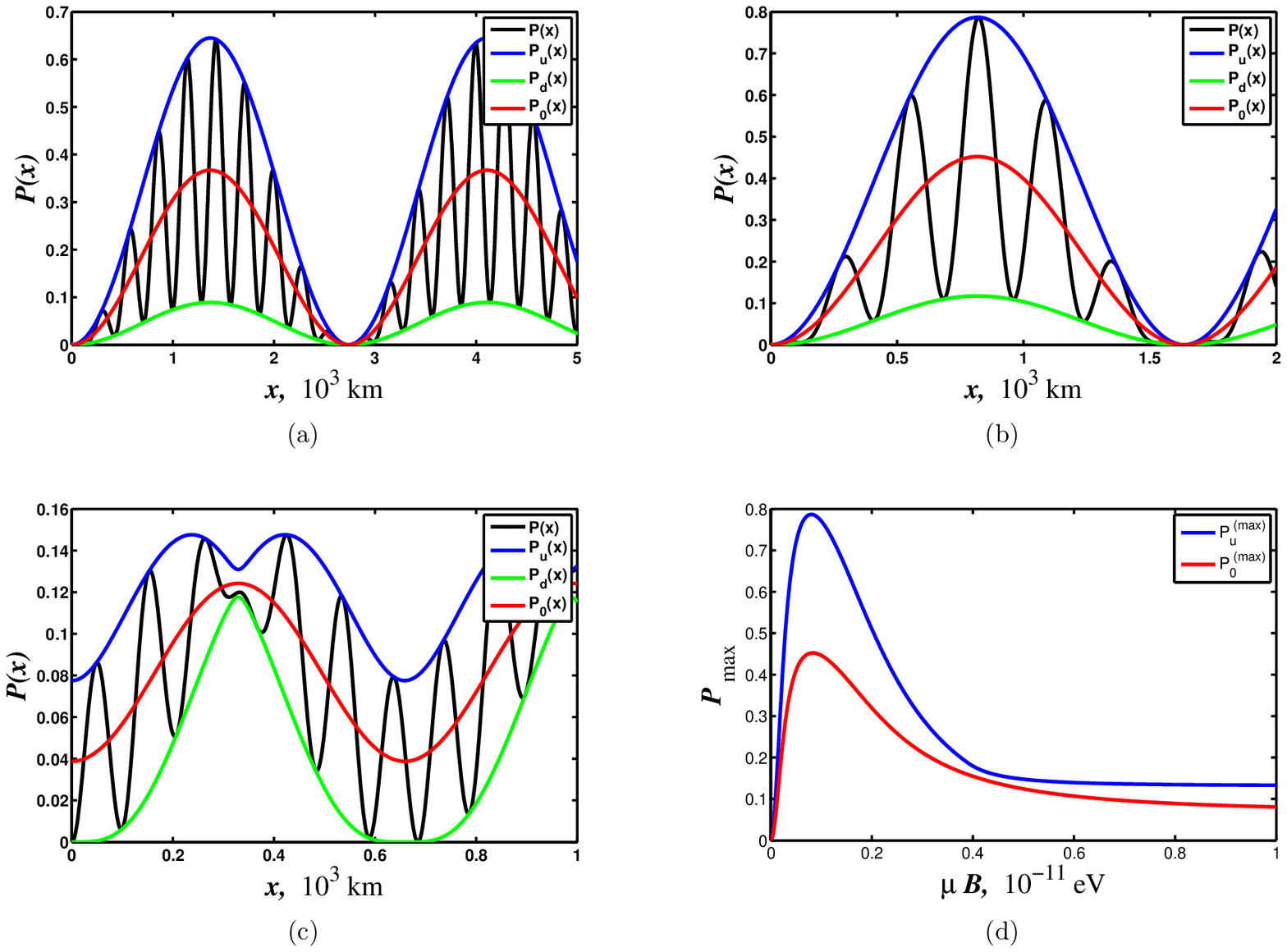}
  \caption{\label{fig3}
  The same as in Fig.~\ref{fig1} for the density
  $\rho = 100\thinspace\text{g/cc}$;
  (a) $\mu B = 4 \times 10^{-13}\thinspace\text{eV}$,
  (b) $\mu B = 8 \times 10^{-13}\thinspace\text{eV}$,
  (c) $\mu B = 5 \times 10^{-12}\thinspace\text{eV}$.}
\end{figure*}

The behavior of the transition probability for various matter
densities $\rho$ and the values of $\mu B$ and for a fixed
neutrino energy of $E=10\thinspace\text{MeV}$ and squared mass
difference of $\delta m^2 = 8 \times 10^{-5}\thinspace\text{eV}^2$
is illustrated in Figs.~\ref{fig1}-\ref{fig3}.

As these plots show, at low matter densities the envelope
functions give, at each propagation distance, the range of the
possible values of the oscillation probability. At greater matter
densities, where the probability oscillates less intensively, the
envelope functions are not that useful in analyzing the physical
situation.

One can find the maximum value of the upper envelope function,
which is also the upper bound for the transition probability,
given as
\begin{widetext}
\begin{equation}\label{PumaxDir}
  P_u^\mathrm{(max)} =
  \begin{cases}
    (S_{+{}}-S_{-{}})^2, & \text{if $B < B'$}, \\
    C_{+{}}C_{-{}}(S_{+{}}^2-S_{-{}}^2)^2
    [C_{+{}}C_{-{}}(S_{+{}}^2+S_{-{}}^2)
    +
    (C_{+{}}C_{-{}})^2+
    (S_{+{}}S_{-{}})^2]^{-1}, &
    \text{if $B > B'$},
  \end{cases}
\end{equation}
\end{widetext}
where the value $B'$ is the solution of the transcendent algebraic
equation, $C_{+{}}C_{-{}}=S_{+{}}S_{-{}}$. The corresponding
maximum values of the averaged transition probablility $P_0(x)$
are given by
\begin{equation}\label{P0maxDir}
  P_0^\mathrm{(max)} =
  \frac{1}{2}[(S_{+{}}S_{-{}})^2-4C_{+{}}C_{-{}}],
\end{equation}
for arbitrary values of $B$. The values of these maxima depend  on
the size of the quantity $\mu B$. These dependencies are plotted
in Figs.~\ref{fig1}(d)-\ref{fig3}(d). In the case of rapid
oscillations the physically relevant quantities, rather than the
maxima, are the averaged values of the transition probability,
which are also plotted in these figures.

As Figs.~\ref{fig1}(d)-\ref{fig3}(d) show, the interplay of the
matter effect and the magnetic interaction can lead, for a given
magnetic moment $\mu$, to an enhanced spin-flavor transition if
the magnetic field $B$ has a suitable strength relative to the
density of matter $\rho$. In our numerical examples this occurs at
$\mu B_\mathrm{max} = 1.1 \times 10^{-12}\thinspace\text{eV}$ for
$\rho = 10\thinspace\text{g/cc}$, at $\mu B_\mathrm{max} = 6.6
\times 10^{-13}\thinspace\text{eV}$ for $\rho =
50\thinspace\text{g/cc}$, and at $\mu B_\mathrm{max} = 8 \times
10^{-13}\thinspace\text{eV}$ for $\rho =
100\thinspace\text{g/cc}$. For these values of $\mu B$ both the
maxima and the average of the transition probability become
considerably larger than for any other values of $\mu B$.
Figures~\ref{fig1}(b)-\ref{fig3}(b) correspond to the situation of
maximal enhancement, whereas Figs.~\ref{fig1}(a)-\ref{fig3}(a) and
Figs.~\ref{fig1}(c)-\ref{fig3}(c) illustrate the situation above
and below the optimal strength $B_\mathrm{max}$ of the magnetic
field.

It is noteworthy that the enhanced transition probability is
achieved towards the lower end of the  $\mu B$ region where
substantial transitions all occur, that is, at relatively moderate
magnetic fields. At larger values of $\mu B$ the maximum of the
transition probability approaches towards $\cos^2(2\theta)$.
Indeed, if $\mu B \gg \max(\Phi,V)$, the transition probability
can be written in the form (see Ref.~\cite{DvoMaa07})
$P(t)=\cos^2(2\theta)\sin^2(\mu B t)$. It was found in
Ref.~\cite{LikStu95} that neutrino spin-flavor oscillations can be
enhanced in a very strong magnetic field, with the transition
probability being practically equal to unity. This phenomenon can
be realized only for Dirac neutrinos with small off-diagonal
magnetic moments and small mixing angle. As we can see from
Figs.~\ref{fig1}(d)-\ref{fig3}(d) the situation is completely
different for big off-diagonal magnetic moments.

One should notice that for long propagation distances consisting
of several oscillation periods of the envelope functions, the
enhancement effect would diminish considerably due to averaging.
In the numerical examples presented in Figs.~\ref{fig1}-\ref{fig3}
the period of the envelope function is of the order of
$10^3-10^4\thinspace\text{km}$, which is a typical size of a shock
wave with the matter densities we have used in the plots (see,
e.g., Ref.~\cite{Kawagoe:2006qu}). Thus the enhanced spin-flavor
transition could take place when neutrinos traverse a shock wave.

Let us recall that the above analysis was made by assuming
neutrinos to be Dirac particles. We will see below (see
Sec.~\ref{MAJNUMATTB}) that the corresponding results are quite
different in the case of Majorana neutrinos.

\section{Evolution of Majorana neutrinos in vacuum}\label{MAJNUVAC}

We now move  to consider Majorana neutrinos, and we shall start
by applying our formalism to the ordinary vacuum oscillation of
two Majorana neutrinos. The left-handed chirality component of a
flavor neutrino $\nu_\lambda^\mathrm{L} =
(1/2)(1-\gamma^5)\nu_\lambda$ is related to the wave functions of
Majorana neutrino states through
\begin{equation}\label{matrtransMaj}
  \nu_\lambda^\mathrm{L} = \sum_a U_{\lambda a}\eta_a,
\end{equation}
where $\lambda=\alpha,\beta$ is the flavor index and $\eta_a$,
$a=1,2$, correspond to a Majorana particle with a definite mass
$m_a$. In the simplest case the mixing of the flavor states arises
purely from Majorana mass terms between the left-handed neutrinos,
and then the mixing matrix $U_{\lambda a}$ is a $2 \times 2$ and
unitary matrix, i.e., $a=1,2$ and, assuming no CP violation, it
can be parametrized in the same way as in
Eq.~\eqref{matrtransDir}.

We study the evolution of this system with the following initial
condition [see also Eq.~\eqref{IniCondDir}]:
\begin{equation}\label{IniCondMaj}
  \nu_\alpha^\mathrm{L}(\mathbf{r},0) = 0,
  \quad
  \nu_\beta^\mathrm{L}(\mathbf{r},0) =
  \nu_\beta^{(0)}e^{\mathrm{i}\mathbf{k}\mathbf{r}},
\end{equation}
where $\mathbf{k}=(0,0,k)$ is the initial momentum and 
$\nu_\beta^{(0)\mathrm{T}}=(0,1)$. The initial state is thus a
left-handed neutrino of flavor $\beta$ propagating along the
$z$-axis to the positive direction.

As both the left-handed state $\nu_\lambda^\mathrm{L}$ and
Majorana state $\eta_a$ have two degrees of freedom, we will
describe them in the following by using two-component Weyl
spinors. The Weyl spinor of a free Majorana particle obeys the
wave equation (see, e.g.,~\cite{FukYan03}),
\begin{equation}\label{WEMajVac}
  \mathrm{i}\dot{\eta}_a+(\bm{\sigma}\mathbf{p}) \eta_a+
  \mathrm{i} m_a\sigma_2\eta_a^{*{}}=0.
\end{equation}
The general solution of this equation can be presented
as~\cite{Cas57}
\begin{align}\label{GenSolMajVac}
  \eta_a(\mathbf{r},t)= &
  \int \frac{\mathrm{d}^3\mathbf{p}}{(2\pi)^{3/2}}
  e^{\mathrm{i}\mathbf{p}\mathbf{r}}
  \sum_{\zeta = \pm 1}
  \big[
    a_a^{(\zeta)}(\mathbf{p})u_a^{(\zeta)}(\mathbf{p}) e^{-\mathrm{i}E_a t}
    \notag
    \\
    & +
    a_a^{(\zeta)*{}}(-\mathbf{p})v_a^{(\zeta)}(-\mathbf{p}) e^{\mathrm{i}E_a t}
  \big],
\end{align}
where $E_a=\sqrt{m_a^2+|\mathbf{p}|^2}$. The basis spinors
$u_a^{(\zeta)}$ and $v_a^{(\zeta)}$ have the form
\begin{alignat}{2}\label{SpinorsMajVac}
  u_a^{+{}}(\mathbf{p}) = & - \lambda_a \frac{m_a}{E_a+|\mathbf{p}|}w_{+{}},
  & \quad
  u_a^{-{}}(\mathbf{p}) = & \lambda_a w_{-{}},
  \\
  \notag
  v_a^{+{}}(\mathbf{p}) = & \lambda_a w_{-{}},
  & \quad
  v_a^{-{}}(\mathbf{p}) = & \lambda_a \frac{m_a}{E_a+|\mathbf{p}|}w_{+{}},
\end{alignat}
where $w_{\pm{}}$ are  helicity amplitudes given
by~\cite{BerLifPit89}
\begin{align}\label{HelAmplMaj}
  w_{+{}} = &
  \begin{pmatrix}
    e^{-i\phi/2}\cos(\vartheta/2) \\
    e^{i\phi/2}\sin(\vartheta/2) \
  \end{pmatrix},
  \notag
  \\
  w_{-{}} = &
  \begin{pmatrix}
    -e^{-i\phi/2}\sin(\vartheta/2) \\
    e^{i\phi/2}\cos(\vartheta/2) \
  \end{pmatrix},
\end{align}
the angles $\phi$ and $\vartheta$ giving the direction of the
momentum of the particle, $\mathbf{p} =
|\mathbf{p}|(\sin\vartheta\cos\phi,\sin\vartheta\sin\phi,\cos\vartheta)$.
The normalization factor $\lambda_a$ in Eq.~\eqref{SpinorsMajVac}
can be chosen as
\begin{equation}\label{NormFactMaj}
  \lambda_a^{-2}=
  1-\frac{m_a^2}{(E_a+|\mathbf{p}|)^2}.
\end{equation}
Let us mention the following  properties of the helicity
amplitudes $w_{\pm{}}$:
\begin{gather}
  (\bm{\sigma}\mathbf{p})w_{\pm{}} = \pm|\mathbf{p}|w_{\pm{}},
  \quad
  \mathrm{i}\sigma_2 w_{\pm{}}^{*{}} = \mp w_{\mp{}},
  \notag
  \\
  w_{\pm{}}(-\mathbf{p}) = \mathrm{i} w_{\mp{}}(\mathbf{p}),
  \notag
  \\
  \left( w_{+{}} \otimes w_{-{}}^\mathrm{T} \right) -
  \left( w_{-{}} \otimes w_{+{}}^\mathrm{T} \right) =
  \mathrm{i}\sigma_2,
  \notag
  \\
  \label{HelAmplProp}
  \left( w_{+{}} \otimes w_{+{}}^\dag \right) +
  \left( w_{-{}} \otimes w_{-{}}^\dag \right) = 1,
\end{gather}
which can be immediately obtained from Eq.~\eqref{HelAmplMaj} and
which are useful in deriving the results given below.

The time-independent coefficients $a_a^{\pm{}}(\mathbf{p})$ in
Eq.~\eqref{GenSolMajVac} have the following form~\cite{Cas57}:
\begin{align}\label{apmMajVac}
  a_a^{+{}}(\mathbf{p}) = & \frac{1}{(2\pi)^{3/2}}
  \bigg[
    \eta_a^{(0)\dag}(-\mathbf{p}) v_a^{+{}}(\mathbf{p})
    \notag
    \\
    & +
    \frac{\mathrm{i}m_a}{E_a+|\mathbf{p}|}
    v_a^{+{}\dag}(-\mathbf{p})\eta_a^{(0)}(\mathbf{p})
  \bigg],
  \notag
  \\
  a_a^{-{}}(\mathbf{p}) = & \frac{1}{(2\pi)^{3/2}}
  \bigg[
    u_a^{-{}\dag}(\mathbf{p}) \eta_a^{(0)}(\mathbf{p})
    \notag
    \\
    & -
    \frac{\mathrm{i}m_a}{E_a+|\mathbf{p}|}
    \eta_a^{(0)\dag}(-\mathbf{p}) u_a^{-{}}(-\mathbf{p})
  \bigg],
\end{align}
where $\eta^{(0)}_a(\mathbf{p})$ is the Fourier transform of the
initial wave function $\eta_a$,
\begin{equation*}
  \eta^{(0)}_a(\mathbf{p}) =
  \int \mathrm{d}^3\mathbf{p}
  e^{-\mathrm{i}\mathbf{p}\mathbf{r}}
  \eta^{(0)}_a(\mathbf{r}).
\end{equation*}
Using Eqs.~\eqref{GenSolMajVac}-\eqref{apmMajVac} we then obtain
the following expression for the wave function for the neutrino
mass eigenstates:
\begin{align}\label{etaaMajVac}
  \eta_a(\mathbf{r},t) = &
  \int \frac{\mathrm{d}^3\mathbf{p}}{(2\pi)^3}
  e^{\mathrm{i}\mathbf{p}\mathbf{r}}
  \lambda_a^2
  \notag
  \\
  & \times
  \Bigg[
    \Bigg\{
      \left(
        e^{-\mathrm{i}E_a t}-
        \left[
          \frac{m_a}{E_a+|\mathbf{p}|}
        \right]^2
        e^{\mathrm{i}E_a t}
      \right)
      \notag
      \\
      & \times
      \left( w_{-{}} \otimes w_{-{}}^\dag \right)
      \notag
      \\
      & +
      \left(
        e^{\mathrm{i}E_a t}-
        \left[
          \frac{m_a}{E_a+|\mathbf{p}|}
        \right]^2
        e^{-\mathrm{i}E_a t}
      \right)
      \notag
      \\
      & \times
      \left( w_{+{}} \otimes w_{+{}}^\dag \right)
    \Bigg\}
    \eta^{(0)}_a(\mathbf{p})
    \notag
    \\ & -
    2\frac{m_a}{E_a+|\mathbf{p}|}
    \sin(E_a t) \sigma_2 \eta^{(0)*{}}_a(-\mathbf{p})
  \Bigg].
\end{align}

From Eqs.~\eqref{HelAmplProp} and~\eqref{etaaMajVac} it follows
that  a mass eigenstate particle initially in the left-polarized
state $\eta_a^{(0)}(\mathbf{r}) \sim
w_{-{}}(\mathbf{k})e^{\mathrm{i}\mathbf{k}\mathbf{r}}$ is
described at later times by
\begin{align}\label{etaaMajVacSC}
  \eta_a(\mathbf{r},t) \sim &
  \lambda_a^2
  \bigg\{
    \left(
      e^{-\mathrm{i}E_a t}-
      \left[
        \frac{m_a}{E_a+|\mathbf{k}|}
      \right]^2
      e^{\mathrm{i}E_a t}
    \right)
    e^{\mathrm{i}\mathbf{k}\mathbf{r}}w_{-{}}(\mathbf{k})
    \notag
    \\ & -
    2\mathrm{i}\frac{m_a}{E_a+|\mathbf{k}|}
    \sin(E_a t) e^{-\mathrm{i}\mathbf{k}\mathbf{r}}w_{+{}}(\mathbf{k})
  \bigg\}.
\end{align}
Let us notice that the second term in Eq.~\eqref{etaaMajVacSC}
describes an antineutrino state. Indeed the spinor
$w_{+{}}(\mathbf{k})$ satisfies the relation,
$(\bm{\sigma}\mathbf{k})w_{+{}}(\mathbf{k}) =
|\mathbf{k}|w_{+{}}(\mathbf{k})$, see Eq.~\eqref{HelAmplProp}.
Therefore it corresponds to an antiparticle, see
Ref.~\cite{BerLifPit89p137}. This term is responsible for the
neutrino-to-antineutrino flavor state transition
$\nu_\beta^\mathrm{L} \leftrightarrow (\nu_\alpha^\mathrm{L})^c$.

According to Eq.~\eqref{matrtransMaj} and~\eqref{matrtransDir},
the wave function of the left-handed neutrino of flavor $\alpha$
is $\nu_\alpha^\mathrm{L} = \cos \theta \eta_1^\mathrm{L} - \sin
\theta \eta_2^\mathrm{L}$. From Eqs.~\eqref{matrtransMaj}
and~\eqref{etaaMajVacSC} it then follows that the probability of
the transition $\nu_\beta^\mathrm{L}\to \nu_\alpha^\mathrm{L}$ in
vacuum is given by
\begin{align}\label{PtrMajVac}
  P_{\nu_\beta^\mathrm{L}\to \nu_\alpha^\mathrm{L}}(t) = &
  |\nu_\alpha^\mathrm{L}|^2 = \sin^2(2\theta)
  \bigg\{
    \sin^2(\Phi t)
    \notag
    \\
    & +
    \frac{1}{4|\mathbf{k}|^2}
    \cos(|\mathbf{k}|t)\sin(\Phi t)
    \notag
    \\
    & \times
    [m_1^2 \sin(E_1 t) - m_2^2 \sin(E_2 t)]
  \bigg\}
  \notag
  \\
  & +
  \mathcal{O}
  \left(
    \frac{m_a}{|\mathbf{k}|}
  \right)^4.
\end{align}
The leading term reproduces the familiar oscillation formula of
Pontecorvo describing the transitions between active neutrinos
$\nu_\beta^\mathrm{L} \leftrightarrow \nu_\alpha^\mathrm{L}$.  The
corrections to Pontecorvo's formula were obtained first in
Ref.~\cite{BlaVit95}, and in our previous
papers~\cite{FOvac,Dvo06EPJC,DvoMaa07,Dvo08} we derived the
analogous corrections for Dirac neutrinos both in vacuum and in
various external fields.

Analogously we can calculate the transition probability for the
process $\nu_\beta^\mathrm{L}\to (\nu_\alpha^\mathrm{L})^c$ using
the second term in Eq.~\eqref{etaaMajVacSC},
\begin{align}\label{PtrnuanuMajVac}
  P_{\nu_\beta^\mathrm{L}\to (\nu_\alpha^\mathrm{L})^c}(t) = &
  |(\nu_\alpha^\mathrm{L})^c|^2 =
  \frac{\sin^2(2\theta)}{4|\mathbf{k}|^2}
  \notag
  \\
  & \times
  [m_1 \sin(E_1 t) - m_2 \sin(E_2 t)]^2
  \notag
  \\
  & +
  \mathcal{O}
  \left(
    \frac{m_a}{|\mathbf{k}|}
  \right)^4.
\end{align}
Note that the next-to-leading term in Eq.~\eqref{PtrMajVac} and
leading term in Eq.~\eqref{PtrnuanuMajVac} have the same order of
magnitude $\sim m_a^2/|\mathbf{k}|^2$.

Before moving to consider Majorana neutrinos in magnetic fields we
make a  general comment concerning the validity of our approach
based on relativistic classical field theory. It has been
stated~\cite{SchVal81} that the dynamics of massive Majorana
fields cannot be described within the classical field theory
approach due to the fact that the mass term of the Lagrangian,
$\eta^\mathrm{T} \mathrm{i}\sigma_2 \eta$, vanishes when $\eta$ is
represented as a $c$-number function. Note that
Eq.~\eqref{WEMajVac} is a direct consequence of the Dirac equation
if we suggest that the four-component wave function satisfies the
Majorana condition. Therefore a solution to Eq.~\eqref{WEMajVac},
i.e., wave functions and energy levels, in principle does not
depend on the existence of a Lagrangian resulting in this
equation. The wave equations describing elementary particles
should follow from the quantum field theory principles. However
quite often these quantum equations allow classical solutions (see
Ref.~\cite{Giu96} for many interesting examples). We have also
demonstrated in Refs.~\cite{FOvac,Dvo06EPJC,DvoMaa07,Dvo08} that
oscillations of Dirac neutrinos in vacuum and various external
fields can be described in the framework of the classical field
theory. The main result of this section was to show that the
quantum Eq.~\eqref{WEMajVac} for massive Majorana particles can be
solved [see Eq.~\eqref{etaaMajVacSC}] in the framework of the
classical field theory as well.

\section{Evolution of Majorana neutrinos in matter and transversal magnetic
field}\label{MAJNUMATTB}

For describing the evolution of two Majorana mass eigenstates in
matter under the influence of an external magnetic field, the wave
Eq.~\eqref{WEMajVac} is to be modified to the following form:
\begin{multline}\label{WEMajmattB}
  \mathrm{i}\dot{\eta}_a +
  \left(
    \bm{\sigma}\mathbf{p}-\frac{g_a}{2}
  \right)\eta_a +
  \mathrm{i} m_a\sigma_2\eta_a^{*{}} - \frac{g}{2}\eta_b
  \\
  -
  \mathrm{i} \mu (\bm{\sigma}\mathbf{B}) \sigma_2 \epsilon_{ab} \eta_b^{*{}}= 0,
  \quad
  a \neq b,
\end{multline}
where $\epsilon_{ab}=\mathrm{i}(\sigma_2)_{ab}$, and $g_a$ and $g$
were defined in connection to Eq.~\eqref{WaveEqDirNu}. Note that
Eq.~\eqref{WEMajmattB} can be formally derived from
Eq.~\eqref{WaveEqDirNu} if one neglects vector current
interactions, i.e., replace $(1-\gamma^5)/2$ with $-\gamma^5/2$,
and takes into account the fact that the magnetic moment matrix of
Majorana neutrinos is antisymmetric (see, e.g.,
Ref.~\cite{FukYan03p477}). We will apply the same initial
condition~\eqref{IniCondMaj} as in the vacuum case. It should be
mentioned that the evolution of Majorana neutrinos in matter and
in a magnetic field has been previously discussed in
Ref.~\cite{Pas96}.

The general solution of Eq.~\eqref{WEMajmattB} can be expressed in
the following form:
\begin{align}\label{GenSolMajmattB}
  \eta_a(\mathbf{r},t)= &
  \int \frac{\mathrm{d}^3\mathbf{p}}{(2\pi)^{3/2}}
  e^{\mathrm{i}\mathbf{p}\mathbf{r}}
  \notag
  \\
  & \times
  \sum_{\zeta = \pm 1}
  \big[
    a_a^{(\zeta)}(\mathbf{p},t)u_a^{(\zeta)}(\mathbf{p})
    \exp(-\mathrm{i}E_a^{(\zeta)} t)
    \notag
    \\
    & +
    a_a^{(\zeta)*{}}(-\mathbf{p},t)v_a^{(\zeta)}(-\mathbf{p})
    \exp(\mathrm{i}E_a^{(\zeta)} t)
  \big],
\end{align}
where the energy levels are given in Eq.~\eqref{EnergyLevQFT} (see
Ref.~\cite{matterQFT}). The basis spinors in
Eq.~\eqref{GenSolMajmattB} can be chosen as
\begin{align*}
  u_a^{+{}}(\mathbf{p}) = & - \lambda_a^{+{}}
  \frac{m_a}{E_a^{+{}}+(|\mathbf{p}|-g_a/2)}w_{+{}},
  \\
  v_a^{-{}}(\mathbf{p}) = & \lambda_a^{-{}}
  \frac{m_a}{E_a^{-{}}+(|\mathbf{p}|+g_a/2)}w_{+{}},
  \\
  u_a^{-{}}(\mathbf{p}) = & \lambda_a^{-{}} w_{-{}},
  \quad
  v_a^{+{}}(\mathbf{p}) = \lambda_a^{+{}} w_{-{}},
\end{align*}
where the normalization factors $\lambda_a^{(\zeta)}$,
$\zeta=\pm{}$  are given by
\begin{equation*}
  (\lambda_a^{(\zeta)})^{-2}=
  1-\frac{m_a^2}{[E_a+(|\mathbf{p}|-\zeta g_a/2)]^2}.
\end{equation*}

Let us consider the propagation of Majorana neutrinos in the
transversal magnetic field. Using a similar technique as in the
Dirac case in Sec.~\ref{DIRNUMATTB} and assuming $k \gg m_a$, we
end up with the following ordinary differential equations for the
coefficients  $a_a^{(\zeta)}$,
\begin{align}\label{Ham1Maj}
  \mathrm{i}\frac{\mathrm{d}\Psi'}{\mathrm{d}t} = & H' \Psi',
  \\
  \notag
  H' = &
  \begin{pmatrix}
    0 & g e^{\mathrm{i} \delta_{-{}} t}/2 & 0 &
    \mu B e^{\mathrm{i} \sigma_{+{}} t} \\
    g e^{-\mathrm{i} \delta_{-{}} t}/2  & 0 &
    -\mu B e^{-\mathrm{i} \sigma_{-{}} t} & 0 \\
    0 & -\mu B e^{\mathrm{i} \sigma_{-{}} t} &
    0 & -g e^{\mathrm{i} \delta_{+{}} t}/2 \\
    \mu B e^{-\mathrm{i} \sigma_{+{}} t} & 0 &
    -g e^{-\mathrm{i} \delta_{+{}} t}/2 & 0 \
  \end{pmatrix},
\end{align}
where
$\Psi^{'\mathrm{T}}=(a_1^{-{}},a_2^{-{}},a_1^{+{}},a_2^{+{}})$ and
\begin{align*}
  \delta_{\pm{}} = & E_1^{\pm{}}-E_2^{\pm{}} \approx 2\Phi \mp \frac{g_1-g_2}{2},
  \\
  \sigma_{\pm{}} = & E_1^{\mp{}}-E_2^{\pm{}} \approx 2\Phi \pm \frac{g_1+g_2}{2}.
\end{align*}

By making the matrix transformation
\begin{widetext}
\begin{equation}
  \Psi' = \mathcal{U}\Psi,
  \quad
  \mathcal{U} = \mathrm{diag}
  \left\{
    e^{\mathrm{i}(\Phi+g_1/2)t},
    e^{-\mathrm{i}(\Phi-g_2/2)t},
    e^{\mathrm{i}(\Phi-g_1/2)t},
    e^{-\mathrm{i}(\Phi+g_2/2)t}
  \right\},
\end{equation}
we can recast Eq.~\eqref{Ham1Maj} into the form
\begin{equation}\label{Ham2Maj}
  \mathrm{i}\frac{\mathrm{d}\Psi}{\mathrm{d}t} = H \Psi,
  \quad
  H =
  \mathcal{U}^\dag H' \mathcal{U} -
  \mathrm{i} \mathcal{U}^\dag \dot{\mathcal{U}}=
  \begin{pmatrix}
    \Phi + g_1/2 & g/2 & 0 & \mu B \\
    g/2 & -\Phi + g_2/2 & -\mu B & 0 \\
    0 & -\mu B & \Phi - g_1/2 & -g/2 \\
    \mu B & 0 & -g/2 & -\Phi - g_2/2 \
  \end{pmatrix}.
\end{equation}
\end{widetext}
Let us note that the analogous effective Hamiltonian has been used
in describing the spin-flavor oscillations of Majorana neutrinos
within the quantum mechanical approach (see, e.g.,
Ref.~\cite{LimMar88}) if we use the basis $\Psi_{QM}^\mathrm{T} =
(\psi_1^\mathrm{L}, \psi_2^\mathrm{L}, [\psi_1^\mathrm{L}]^c,
[\psi_2^\mathrm{L}]^c)$.

Note that the consistent derivation of the master
Eq.~\eqref{WEMajmattB} should be done in the framework of the
quantum field theory (see, e.g., Ref.~\cite{SchVal81}), supposing
that the spinors $\eta_a$ are expressed via anticommuting
operators. This quantum field theory treatment is important to
explain the asymmetry of the magnetic moment matrix. However, it
is possible to see that the main Eq.~\eqref{WEMajmattB} can also
be reduced to the standard Schr{\"o}dinger evolution
Eq.~\eqref{Ham2Maj} for neutrino spin-flavor oscillations if we
suppose that the wave functions $\eta_a$ are $c$-number objects.
That is why one can again conclude that classical and quantum
field theory methods for studying Majorana neutrinos' propagation
in external fields are equivalent.

Let us again consider the situation when $n_e = n_p = n_n = n$,
which results in $g_1 = -g_2$. In this case the eigenvalues of the
Hamiltonian~\eqref{Ham2Maj} $\lambda = \pm \mathcal{E}_{\pm{}}$
are given by
\begin{align}
  \mathcal{E}_{\pm{}}= & \frac{1}{2}
  \sqrt{V^2+4(\mu B)^2+4\Phi^2 \pm 4VR},
  \notag
  \\
  R = & \sqrt{(\Phi\cos 2\theta)^2+(\mu B)^2},
\end{align}
where $V = G_\mathrm{F} n/\sqrt{2}$ as in Sec~\ref{DIRNUMATTB}.
The time evolution of the wave function is described by the
formula,
\begin{align}\label{PsimattBMaj}
  \Psi(t)= & \sum_{\zeta = \pm 1}
  \Big[
    \left(
      U_\zeta \otimes U_\zeta^\dag
    \right)\exp{(-\mathrm{i}\mathcal{E}_\zeta t)}
    \notag
    \\
    & +
    \left(
      V_\zeta \otimes V_\zeta^\dag
    \right)\exp{(\mathrm{i}\mathcal{E}_\zeta t)}
  \Big]\Psi_0,
\end{align}
where  $U_\zeta$ and $V_\zeta$ are the eigenvectors of the
Hamiltonian~\eqref{Ham2Maj}, given as
\begin{equation}\label{BasisSpinorsMaj}
  U_\zeta =
  \frac{1}{N_\zeta}
  \begin{pmatrix}
    -x_\zeta \\
    -y_\zeta \\
    1 \\
    -z_\zeta \
  \end{pmatrix},
  \quad
  V_\zeta =
  \frac{1}{N_\zeta}
  \begin{pmatrix}
    -y_\zeta \\
    x_\zeta \\
    z_\zeta \\
    1 \
  \end{pmatrix},
\end{equation}
where
\begin{align}\label{xyzSigma}
  x_\zeta = & \frac{\mu B (\mathcal{E}_\zeta+\Phi)}{\Sigma_\zeta} V \sin 2\theta,
  \notag
  \\
  y_\zeta = & \frac{\mu B}{\mathcal{E}_\zeta+\Phi-V \cos 2\theta /2}
  \left[
    1+\frac{(\mathcal{E}_\zeta+\Phi)}{2 \Sigma_\zeta} V^2 \sin^2(2\theta)
  \right],
  \notag
  \\
  z_\zeta = & \frac{V\sin 2\theta}{2(\mathcal{E}_\zeta+\Phi+V \cos 2\theta /2)}
  \left[
    1+\frac{2(\mu B)^2(\mathcal{E}_\zeta+\Phi)}{\Sigma_\zeta}
  \right],
  \notag
  \\
  \Sigma_\zeta = & \frac{V}{2}
  [2\mathcal{E}_\zeta(\mathcal{E}_\zeta+\Phi)-V^2/2 + \Phi V \cos 2\theta]
  \cos 2\theta
  \notag
  \\
  & +
  \zeta R V
  (\mathcal{E}_\zeta+\Phi-V \cos 2\theta /2).
\end{align}
The normalization coefficient $N_\zeta$ in
Eq.~\eqref{BasisSpinorsMaj} is given by
$N_\zeta=\sqrt{1+x_\zeta^2+y_\zeta^2+z_\zeta^2}$.

Proceeding along the same lines as in Sec.~\ref{DIRNUMATTB}, we
obtain from Eqs.~\eqref{matrtransMaj}
and~\eqref{PsimattBMaj}-\eqref{xyzSigma}  the probability of the
process $\nu_\beta^\mathrm{L} \to \nu_\alpha^\mathrm{R}$ as,
\begin{align}\label{PtrmattBMaj}
  P_{\nu_\beta^\mathrm{L} \to \nu_\alpha^\mathrm{R}}(t) = &
  [C_{+{}}\cos(\mathcal{E}_{+{}}t)+C_{-{}}\cos(\mathcal{E}_{-{}}t)]^2
  \notag
  \\
  & +
  [S_{+{}}\sin(\mathcal{E}_{+{}}t)+S_{-{}}\sin(\mathcal{E}_{-{}}t)]^2,
\end{align}
where
\begin{align}\label{CpmSpmMaj}
  C_\zeta = & -\frac{1}{N_\zeta^2}[\sin 2\theta (x_\zeta + y_\zeta z_\zeta) +
  \cos 2\theta (y_\zeta - x_\zeta z_\zeta)],
  \notag
  \\
  S_\zeta = & \frac{1}{N_\zeta^2}(y_\zeta + x_\zeta z_\zeta).
\end{align}
Consistently with Eq.~\eqref{IniCondMaj}, we have taken the
initial  wave function as
\begin{equation}\label{IniCondQMMaj}
  \Psi_0^\mathrm{T}=(\sin\theta,\cos\theta,0,0).
\end{equation}
With help of Eqs.~\eqref{xyzSigma} and~\eqref{CpmSpmMaj} it is
easy to check that $C_{+{}}+C_{-{}}=0$  guaranteeing $P(0)=0$.

Note that formally Eq.~\eqref{PtrmattBMaj} corresponds to the
transitions $\nu_\beta^\mathrm{L} \to \nu_\alpha^\mathrm{R}$.
However, virtually it describes oscillations between active
neutrinos $\nu_\beta^\mathrm{L} \leftrightarrow
(\nu_\alpha^\mathrm{L})^c$ since
$\nu_\alpha^\mathrm{R}=(\nu_\alpha^\mathrm{L})^c$ for Majorana
particles.

\begin{figure*}
  \centering
  \includegraphics[scale=.93]{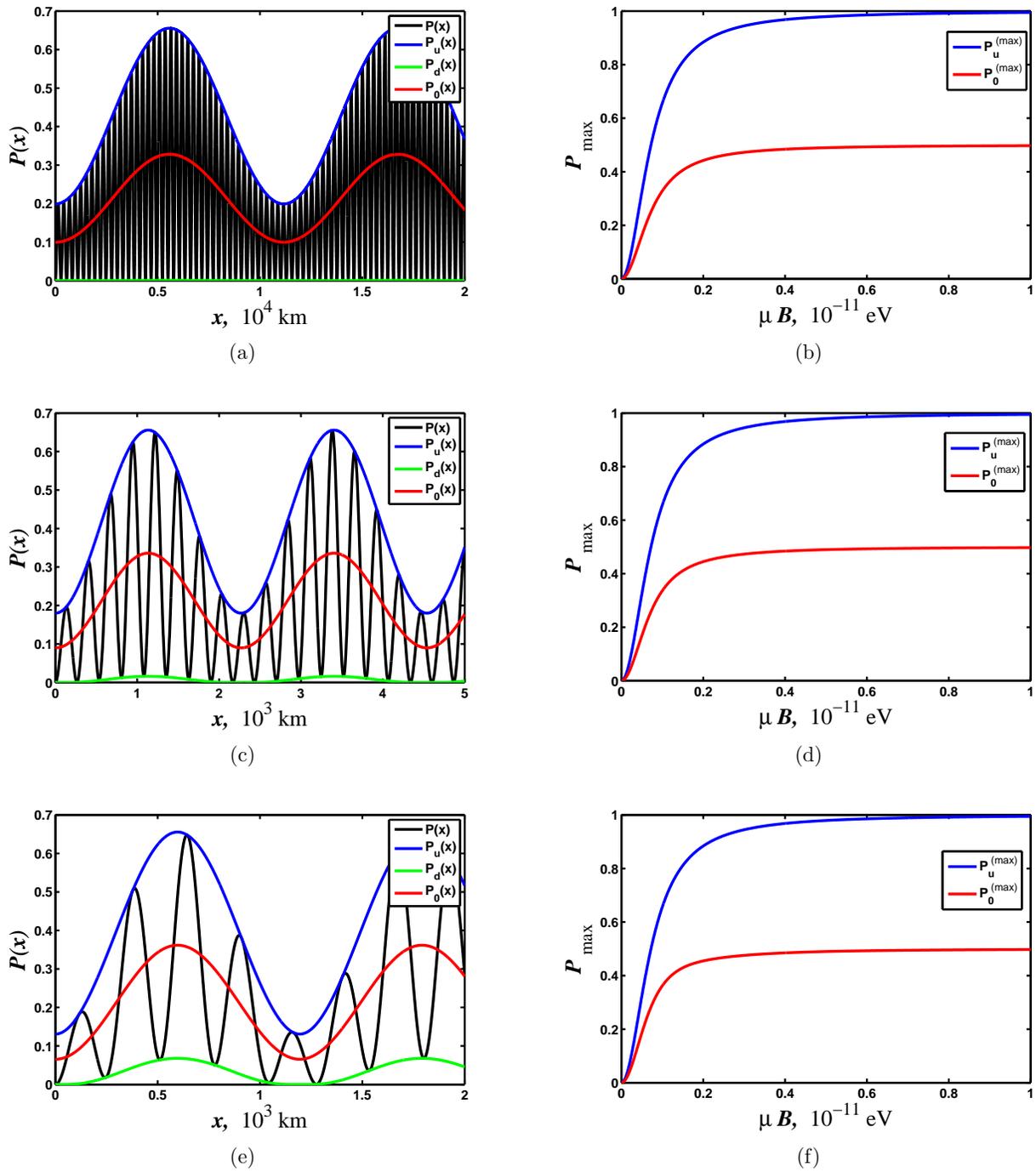}
  \caption{\label{fig4}
  (a), (c) and (e) The transition probability versus the
  distance passed by a neutrino beam. The neutrino parameters
  have the same values as in Figs.~\ref{fig1}-\ref{fig3},
  $E_\nu = 10\thinspace\text{MeV}$,
  $\delta m^2 = 8 \times 10^{-5}\thinspace\text{eV}^2$ and
  $\theta = 0.6$. The magnetic energy is equal to
  $\mu B = 10^{-12}\thinspace\text{eV}$.
  The black line is the function $P(x)$,
  the blue and green lines are the envelope functions $P_{u,d}(x)$ and
  the red line is the averaged transition probability $P_0(x)$.
  (b), (d) and (f)
  The dependence of the maximal values of the functions $P(x)$ and
  $P_0(x)$, blue and red lines respectively, on the magnetic energy
  for the given density.
  Panels (a) and (b) correspond to the matter density
  $\rho = 10\thinspace\text{g/cc}$,
  (c) and (d) -- to $\rho = 50\thinspace\text{g/cc}$ and
  (e) and (f) -- to $\rho = 100\thinspace\text{g/cc}$.}
\end{figure*}

As in the previous case of Eq.~\eqref{PrtDir},
Eq.~\eqref{PtrmattBMaj} can be treated  analytically for
relatively small values of the effective potential $V$. The
ensuing envelope functions $P_{u,d} = P_0 \pm \sqrt{P_c^2+P_s^2}$
depend on the coefficients $C_\zeta$ and $S_\zeta$ in the same way
as in Eq.~\eqref{P0PcPsDir}. The  transition probabilities at
various  values of the matter density and the magnetic field are
presented in Fig.~\ref{fig4}.

Despite the formal similarity between Dirac and Majorana
transition probabilities (see Eqs.~\eqref{CpmSpmDir}
and~\eqref{CpmSpmMaj}) the actual dynamics is quite different in
these two cases, as one can see by comparing
Figs.~\ref{fig1}(d)-\ref{fig3}(d) and Fig.~\ref{fig4}, panels (b),
(d) and (f). In particular, in the Majorana case
$P_u^\mathrm{(max)}=4|C_{+{}}C_{-{}}|$ for arbitrary $B$, to be
compared with Eq.~\eqref{PumaxDir}, while the function
$P_0^\mathrm{(max)}$ has the same form as in the Dirac case given
in Eq.~\eqref{P0maxDir}. In contrast with the Dirac case, the
averaged transition probability does not achieve its maximal value
at some moderate magnetic field $B_\mathrm{max}$ value, but both
$P_u^\mathrm{max}$ and $P_0^\mathrm{max} $ are monotonically
increasing functions of the strength of the magnetic field with 1
and 1/2 as their asymptotic values, respectively. We can
understand this behavior when we recall that, at $\mu B \gg
\max(\Phi,V)$, the effective Hamiltonian  Eq.~\eqref{Ham2Maj}
becomes
\begin{equation}\label{HinfMaj}
  H_\infty =
  \mathrm{i} \mu B \gamma^2,
  \quad
  \mathrm{i}\gamma^2 =
  \begin{pmatrix}
    0 & 0 & 0 & 1 \\
    0 & 0 & -1 & 0 \\
    0 & -1 & 0 & 0 \\
    1 & 0 & 0 & 0 \
  \end{pmatrix}.
\end{equation}
The Schr\"odinger equation with the effective
Hamiltonian~\eqref{HinfMaj} has the formal solution
\begin{align}\label{FormSolMaj}
  \Psi(t)= & \exp(-\mathrm{i} H_\infty t)\Psi(0)
  \notag
  \\
  = &
  [\cos(\mu B t)+\gamma^2\sin(\mu B t)]\Psi(0).
\end{align}
Using Eqs.~\eqref{matrtransMaj}, \eqref{IniCondQMMaj}
and~\eqref{FormSolMaj} we then immediately arrive to the following
expression for the transition probability,
$P(t)=|\nu_\alpha^\mathrm{R}|^2=\sin^2(\mu B t)$, which explains
the behavior of the function $P_u^\mathrm{(max)}$ at strong
magnetic fields. Note that the analogous result was also obtained
in Ref.~\cite{LikStu95}.

Finally, it is worth of noticing that in contrast to the Dirac
case, the behavior of the transition probability in the Majorana
case is qualitatively similar for different matter densities and
different magnetic fields [see Fig.~\ref{fig4}, panels (a), (c),
and (e)].

The problem of Majorana neutrinos' spin-flavor oscillations was
studied in Refs.~\cite{AndSatPRD03,AndSatJCAP03} with help of
numerical codes. For example, in Ref.~\cite{AndSatPRD03} the
evolution equation for three neutrino flavors propagating inside a
presupernova star with zero metallicity, e.g., corresponding to
W02Z model~\cite{Woosley:2002zz}, was solved for the realistic
matter and magnetic field profiles. Although our analytical
transition probability formula~\eqref{PtrmattBMaj} is valid only
for the constant matter density and magnetic field strength, it is
interesting to compare our results with the numerical simulations
of Ref.~\cite{AndSatPRD03}. In those calculations the authors used
magnetic fields $B \sim 10^{10}\thinspace\text{G}$ and magnetic
moments $\sim 10^{-12}\thinspace\mu_\mathrm{B}$ that give us the
magnetic energy $\mu B \sim 10^{-11}\thinspace\text{eV}$. This
value is the maximal magnetic energy used in our work.

It was found in Ref.~\cite{AndSatPRD03} that spin-flavor
conversion is practically adiabatic for low-energy neutrinos
corresponding to $E_\nu \sim 5\thinspace\text{MeV}$ inside the
region where $Y_e \approx 0.5$ and the averaged transition
probability for the channel $\nu_\mu \to \bar{\nu}_e$ is close
$0.5$. This big transition probability is due to the RSF-H and
RSF-L resonances at the distance $\approx 0.01 R_\odot$. Even so
that we study two neutrino oscillations scheme, we obtained the
analogous behavior of $P_0^\mathrm{(max)}$ (see Fig.~\ref{fig4}).
However, in our case this big transition probability is due to the
presence of the strong magnetic field (see
Refs.~\cite{GiuStu08,LikStu95}). We cannot compare our transition
probability formula~\eqref{PtrmattBMaj} with the results of
Ref.~\cite{AndSatPRD03} for higher energies, $E_\nu
> 25\thinspace\text{MeV}$, since spin-flavor oscillations become
strongly nonadiabatic for these kinds of energies and one has to
take into account the coordinate dependence of the matter density
which should decrease with radius as $1/r^3$~\cite{Kac02}.


\section{Summary}\label{CONCL}

We have studied in this work the propagation of massive
flavor-mixed Dirac and Majorana particles in matter under the
influence of an external magnetic field in an approach based on
relativistic quantum mechanics. The magnetic moment matrix   is
assumed to be both in the Dirac and Majorana case nondiagonal in
the mass eigenstate basis. Starting with  Lorentz invariant wave
equations, the interactions of neutrinos with matter and magnetic
fields included, we derived the effective Hamiltonians and solve
the evolution equations for the electrically neutral matter with
$n_n = n_p$. We then found the  probabilities for the spin-flavor
oscillation  $\nu_\beta^\mathrm{L} \to \nu_\alpha^\mathrm{R}$
($\nu_\beta^\mathrm{L} \to (\nu_\alpha^\mathrm{L})^c$ for Majorana
neutrinos) and examined their dependence  on the matter density
and the strength of the magnetic field.

We also studied in our approach the evolution of mixed massive
Majorana neutrinos in vacuum, which is an exactly solvable
problem. We derived the  wave functions of the neutrinos and the
transition probabilities for the process $\nu_\beta \to
\nu_\alpha$. We included the terms quadratic in neutrino masses
and discussed the origin of these corrections of the standard
formula for the probability.

These corrections to the Pontecorvo formula, see
Eq.~\eqref{PtrMajVac}, are rapidly oscillating functions
suppressed by the ratio $m_a/|\mathbf{k}|$, which is small for the
relativistic neutrinos. It was revealed in our previous
works~\cite{FOvac,Dvo06EPJC,DvoMaa07,Dvo08} that the analogous
terms originate from the accurate account of the Lorentz
invariance in the study of oscillations of Dirac neutrinos. It is
also known that in describing of evolution of Majorana neutrinos
in vacuum the transitions of the types $\nu_\beta^\mathrm{L} \to
\nu_\alpha^\mathrm{L}$ and $\nu_\beta^\mathrm{L} \to
(\nu_\alpha^\mathrm{L})^c$ are possible~\cite{Kob82}, with the
transition probability of the latter case being suppressed by the
factor $m_a/|\mathbf{k}|$. The transitions $\nu_\beta^\mathrm{L}
\to (\nu_\alpha^\mathrm{L})^c$ can be interpreted as
neutrino-to-antineutrino oscillations and manifest in the possible
neutrinoless double beta decay~\cite{Bar02}. In our work on the
basis of the relativistic quantum mechanics we derived the
transition probability for neutrino-to-antineutrino oscillations
which is consistent with the results of Ref.~\cite{Kob82}.

Our results can be used for analyzing the behavior of neutrinos in
supernovae, in particular in the so-called zero metallicity
presupernovae,  where the condition  $n_n = n_p$ is
fulfilled~\cite{Woosley:2002zz} outside the Si+O layer. It is
known that these stars can possess very strong magnetic fields, up
to $10^{15}\thinspace\text{G}$, or even stronger~\cite{PriRos06}.

We found that large rates of transitions that change neutrino
flavors and chiralities are possible both in the Dirac and
Majorana cases with a typical oscillation lengths of the order of
$10^3-10^4\thinspace\text{km}$ for matter densities of the order
of $10-100\thinspace\text{g/cc}$, for typical supernova neutrino
energies of $10\thinspace\text{MeV}$ and for $\mu B$ of the order
of $10^{-13} - 10^{-12}\thinspace\text{eV}$. The parameters of a
shock wave, the  density and width, can vary in a rather wide
range, but typically they fit with the parameter values we used in
our analysis~\cite{supernovaparam}. In particular, the width of
the shock wave can be of about the same size as the oscillation
length of the spin-flavor oscillations. It would thus be possible
that the transition probability achieves its maximum value (can be
as high as 0.8 in some cases) when neutrinos pass through the
shock wave on their way towards the outer layers of the star.  If
the flight distance were many oscillation lengths, the probability
would be averaged to a smaller value. The results of our work can
have the implication to the r-process nucleosynthesis since it is
sensitive to the amount of neutrinos of various species emitted in
a supernova~\cite{BalFul03}.

In the Dirac case the maximum value of the transition probability
is achieved for a specific value of $\mu B$, while in the Majorana
case the transition probability can be large for a wide range of
$\mu B$ values. Supposing that the neutrino magnetic moment is $3
\times 10^{-12}\mu_\mathrm{B}$, allowed by the present
astrophysical and cosmological data~\cite{Raf90}, the largest
transition in the Dirac case would occur when the strength of the
magnetic field is in the range $10^8 - 10^9 \thinspace\text{G}$.
Although a magnetic field on the surface of a neutron star is
typically stronger, $10^{12} - 10^{13} \thinspace\text{G}$, at
outer parts of the envelope it may have a suitable  value for a
large transition to take place. The difference between the
dynamics of Dirac and Majorana particles under the influence of
the same external fields can be a smoking gun to reveal the nature
of neutrinos.

Note that in the majority of the previous works devoted to the
spin-flavor oscillations the analytical transition probability
formulae were obtained only for Majorana neutrinos. The case of
Dirac particles was studied only in connection with small
off-diagonal magnetic moments. Our work studies the opposite
situation of big off-diagonal elements of the magnetic moments
matrix. The effect of the appearance of the big transition
probability at moderate magnetic field strength has never been
described earlier. As for Majorana neutrinos the enhancement of
the transition probability in strong magnetic fields is in
agreement with the previous studies~\cite{GiuStu08,LikStu95}. Note
that spin-flavor oscillations of Majorana neutrinos can be also
resonantly enhanced in the moderate magnetic field, see
Ref.~\cite{OscSolarNu}. However this effect does not happen in
isoscalar matter with $n_e=n_p=n_n$. The condition for the
resonant spin-flavor precession reads~\cite{OscSNNu},
$G_\mathrm{F}(n_e-n_n) = \pm \delta m^2 \cos 2\theta/4k$, which is
not fulfilled in isoscalar matter.

The obtained analytical formulae for the transition probabilities,
Eqs.~\eqref{PrtDir} and~\eqref{PtrmattBMaj}, are valid in the
important case when $Y_e = 0.5$, which is realized in the zero
metallicity presupernova stars. These kinds of stars were quite
common for early stages of the galaxies formation. We obtained
that spin-flavor oscillations of Dirac neutrinos can be strongly
influenced by the moderate magnetic field of a neutron star and
the big transition probability can exist not only for Majorana
neutrinos~\cite{AndSatPRD03,AndSatJCAP03}. Thus spin-flavor
oscillations of Dirac neutrinos can significantly change the relic
supernova neutrino background~\cite{relicNU}. Although now no
signal of relic neutrinos was observed~\cite{Mal03}, there are
still some efforts to calculate the flux of relic $\bar{\nu}_e$
for the KamLAND and Super-Kamiokande detectors~\cite{Str04}. The
reliable simulation of the supernova explosion does not exist yet.
For example, the propagation of a shock wave can significantly
change magnitude and shape of the magnetic field which are very
important in our calculations. Nevertheless it is believed that
the future galactic supernova neutrino burst might give some clues
to the physics of relic supernova neutrinos~\cite{AndSatJCAP03}.

\begin{acknowledgments}
The work has been supported by the Academy of Finland under
Contracts Nos.~108875 and~104915. MD is thankful to the Conicyt
(Chile), Programa Bicentenario PSD-91-2006, for the grant as well
as to Victor Semikoz (IZMIRAN) for helpful discussions and to the
anonymous referee for valuable comments.
\end{acknowledgments}


\begin{thebibliography}{100}

\bibitem{DiracBosc}
  A.~Cisneros,
  Astrophys. Space Sci. \textbf{10}, 87 (1971);
  M.~B.~Voloshin, M.~I.~Vysotski\u{\i}, and L.~B.~Okun',
  Sov. Phys. JETP \textbf{64}, 446 (1986).

\bibitem{SchVal81}
  J.~Schechter and J.~W.~F. Valle,
  Phys. Rev. D \textbf{24}, 1883 (1981).

\bibitem{LimMar88}
  C.-S.~Lim and W.~J.~Marciano,
  Phys. Rev. D \textbf{37}, 1368 (1988).

\bibitem{Akh88}
  E.~Kh.~Akhmedov,
  Sov. J. Nucl. Phys. \textbf{48}, 382 (1988).

\bibitem{OscSolarNu}
  J.~Pulido,
  Phys. Rep. \textbf{211}, 167 (1992);
  E.~Kh.~Akhmedov, S.~T.~Petcov, and A.~Yu.~Smirnov,
  Phys. Rev. D \textbf{48}, 2167 (1993), hep-ph/9301211;
  A.~B.~Balantekin and P.~Loreti,
  Phys. Rev. D \textbf{48}, 5496 (1993);
  J.~Barranco \textit{et~al}.,
  Phys. Rev. D \textbf{66}, 093009 (2002), hep-ph/0207326;
  A.~Friedland and A.~Gruzinov,
  Astropart. Phys. \textbf{19}, 575 (2003), hep-ph/0202095;
  M.~Picariello, \textit{et~al.},
  JHEP \textbf{11}, 055 (2007), 0705.4070 [hep-ph].

\bibitem{smallcontr}
  E.~Kh.~Akhmedov and J.~Pulido,
  Phys. Lett. B \textbf{553}, 7 (2003),
  hep-ph/0209192;
  A.~B.~Balantekin and C.~Volpe,
  Phys. Rev. D \textbf{72}, 033008 (2005),
  hep-ph/0411148.

\bibitem{OscSNNu}
  H.~Nunokawa, Y.-Z.~Qian, and G.~M.~Fuller,
  Phys. Rev. D \textbf{55}, 3265 (1997), astro-ph/9610209;
  H.~Nunokawa, R.~Tomas, and J.~W.~F.~Valle,
  Astropart. Phys. \textbf{11}, 317 (1999), astro-ph/9811181;
  E.~Kh.~Akhmedov and T.~Fukuyama,
  JCAP \textbf{0312}, 007 (2003), hep-ph/0310119.

\bibitem{AndSatPRD03}
  S.~Ando and K.~Sato,
  Phys. Rev. D \textbf{68}, 023003 (2003), hep-ph/0305052.

\bibitem{AndSatJCAP03}
  S.~Ando and K.~Sato,
  JCAP \textbf{0310}, 001 (2003), hep-ph/0309060.

\bibitem{GiuStu08}
  C.~Giunti and A.~Studenikin,
  0812.3646 [hep-ph] (2008).

\bibitem{FOvac}
  M. Dvornikov,
  Phys. Lett. B \textbf{610}, 262 (2005), hep-ph/0411101;
  in
  \textit{Proceedings of the IPM school and conference on Lepton and
  Hadron Physics,
  Tehran, 2006}, ed. by Y.~Farzan, eConf~C0605151 (2007),
  hep-ph/0609139;
  Phys. Atom. Nucl. \textbf{72}, 116 (2009), hep-ph/0610047.

\bibitem{Dvo06EPJC}
  M. Dvornikov,
  Eur. Phys. J. C \textbf{47}, 437 (2006), hep-ph/0601156;
  J. Phys. Conf. Ser. \textbf{110}, 082005 (2008), 0708.2975 [hep-ph].

\bibitem{DvoMaa07}
  M.~Dvornikov and J.~Maalampi, Phys. Lett. B \textbf{657}, 217 (2007);
  hep-ph/0701209.

\bibitem{Dvo08}
  M.~Dvornikov, J. Phys. G \textbf{35}, 025003 (2008); 0708.2328 [hep-ph].

\bibitem{EllEng04}
  S.~R. Elliott and J.~Engel,
  J. Phys. G \textbf{30}, R183 (2004), hep-ph/0405078.

\bibitem{Giunti:2007ry}
  C.~Giunti and C.~W.~Kim,
  \textit{Fundamentals of Neutrino Physics and Astrophysics}
  (Oxford Univ. Press, Oxford, 2007),
  pp.~320--321.

\bibitem{matteffpot}
  L.~Wolfenstein, Phys. Rev. D \textbf{17}, 2369 (1978);
  M.~Dvornikov and A.~Studenikin,
  JHEP \textbf{09}, 016 (2002), hep-ph/0202113.

\bibitem{Bel05}
  N.~F.~Bell, \textit{et al.},
  Phys. Rev. Lett. \textbf{95}, 151802 (2005); hep-ph/0504134.



\bibitem{ArbLobMur07}
  E.~V.~Arbuzova, A.~E.~Lobanov, and E.~M.~Murchikova,
  Phys. Atom. Nucl. \textbf{72}, 141 (2009), 0711.2649 [hep-ph].

\bibitem{matterQFT}
  A.~Studenikin and A.~Ternov,
  Phys. Lett. B \textbf{608}, 107 (2005), hep-ph/0412408;
  A.~E.~Lobanov,
  Phys. Lett. B \textbf{619}, 136 (2005), hep-ph/0506007.

 \bibitem{Woosley:2002zz}
  S.~E.~Woosley, A.~Heger and T.~A.~Weaver,
  Rev. Mod. Phys. \textbf{74}, 1015 (2002).

\bibitem{LikStu95}
  G.~G.~Likhachev and A.~I.~Studenikin,
  JETP \textbf{81}, 419 (1995).

\bibitem{Kawagoe:2006qu}
  S.~Kawagoe, \textit{et al.},
  J. Phys. Conf. Ser. \textbf{39}, 294 (2006).

\bibitem{FukYan03}
  M.~Fukugita and T.~Yanagida,
  \textit{Physics of Neutrinos and Applications to Astrophysics}
  (Springer, Berlin, 2003),
  pp.~292--296.

\bibitem{Cas57}
  K.~M.~Case,
  Phys. Rev. \textbf{107}, 307 (1957).

\bibitem{BerLifPit89}
  V.~B.~Berestetski\u{\i}, E.~M.~Lifschitz, and L.~P.~Pitaevski\u{\i},
  \textit{Quantum Electrodynamics}
  (Moscow, Nauka, 1989),
  pp.~108--112.

\bibitem{BerLifPit89p137}
  See pp.~137--141 in Ref.~\cite{BerLifPit89}.

\bibitem{BlaVit95}
  M.~Blasone and G.~Vitiello,
  Ann. Phys. \textbf{244}, 283 (1995),
  Erratum \textit{ibid.} \textbf{249} (1996) 363,
  hep-ph/9501263.

\bibitem{Giu96}
  D. Giulini, \textit{et al.},
  \textit{Decoherence and the Appearence of a Classical
  World in Quantum Theory}
  (Springer-Verlag, Berlin, 1996).

\bibitem{FukYan03p477}
  See pp.~477--478 in Ref.~\cite{FukYan03}.

\bibitem{Pas96}
  S.~Pastor, \textit{Master's Thesis} (University of Valencia, 1996).

\bibitem{Kac02}
  M.~Kachelrie\ss, \textit{et al.},
  Phys. Rev. D \textbf{65}, 073016 (2002), hep-ph/0108100.

\bibitem{Kob82}
  I.~Yu.~Kobzarev, \textit{et al.},
  Sov. J. Nucl. Phys. \textbf{35}, 708 (1982).

\bibitem{Bar02}
  G.~Barenboim, \textit{et al.},
  Phys. Lett. B \textbf{537}, 227 (2002), hep-ph/0203261.

\bibitem{PriRos06}
  D.~J.~Price and S.~Rosswog,
  Science \textbf{312}, 719 (2006).

\bibitem{supernovaparam}
  A.~Burrows, K.~Klein, and R.~Gandhi,
  Phys. Rev. D \textbf{45}, 3361 (1992);
  R.~Tom\`as, \textit{et~al.},
  JCAP \textbf{0409}, 015 (2004), astro-ph/0407132.

\bibitem{BalFul03}
  A.~B.~Balantekin and G.~M.~Fuller,
  J. Phys. G \textbf{29}, 2513 (2003), astro-ph/0309519.

\bibitem{Raf90}
  G.~G.~Raffelt,
  Phys. Rev. Lett. \textbf{64}, 2856 (1990).

\bibitem{relicNU}
  See pp.~515--516 in Ref.~\cite{Giunti:2007ry}.

\bibitem{Mal03}
  M.~Malek, \textit{et al.},
  Phys. Rev. Lett. \textbf{90}, 061101 (2003), hep-ex/0209028.

\bibitem{Str04}
  L.~E.~Strigari, \textit{et al.},
  JCAP \textbf{0403}, 007 (2004), astro-ph/0312346.

\end{thebibliography}
\end{document}